%% file: Paper_3D_IFAC_WC.tex
\documentclass{ifacconf}

\usepackage[utf8]{inputenc}
\usepackage{amsmath,amssymb}
\usepackage{booktabs}
\usepackage{multirow}
\usepackage{grffile}
\usepackage{standalone}
\usepackage[table]{xcolor}
\usepackage{tikz}
\usetikzlibrary{arrows}
\usepackage{bm}
\usepackage{xstring}
\usetikzlibrary{calc}
\usepackage{subfigure}
\usepackage{pgfplots}
\usepackage{import}
\usepackage{xifthen}
\usepackage{tikz-3dplot}
\usetikzlibrary{plotmarks}
\usetikzlibrary{arrows.meta}
\usetikzlibrary{positioning, decorations}
\definecolor{link_red}{RGB}{196, 19, 46}
\definecolor{delta_color}{rgb}{1.0000,0.4941,0.0510}

\usepackage{xparse,mathtools}
\usepackage{adjustbox}

\makeatletter
\let\old@ssect\@ssect % Store how ifacconf defines \@ssect
\makeatother

\usepackage{natbib}
\usepackage{hyperref}

\makeatletter
\def\@ssect#1#2#3#4#5#6{%
  \NR@gettitle{#6}% Insert key \nameref title grab
  \old@ssect{#1}{#2}{#3}{#4}{#5}{#6}% Restore ifacconf's \@ssect
}
\makeatother

\usepgfplotslibrary{external}
\immediate\write18{if not exist tikz_external mkdir tikz_external}
\tikzset{
    export as png/.style={
        external/system call/.add={}{
            && convert -density #1 -transparent white "\image.pdf" "\image.png"
        },
    },
    export as png/.default={200},
}

\input{./macros.tex}
\begin{document}
\begin{frontmatter}

\title{Magnetometer-free inertial motion tracking of arbitrary joints with range of motion constraints}

\author[First]{Dustin Lehmann}
\author[Second]{Daniel Laidig}
\author[Third]{Raphael Deimel}
\author[Fourth]{Thomas Seel}

\address[First]{Technische Universität Berlin, Control Systems Group (e-mail: dustin.lehmann@tu-berlin.de).}
\address[Second]{Technische Universität Berlin, Control Systems Group (e-mail: laidig@control.tu-berlin.de)}
\address[Third]{Technische Universität Berlin (e-mail: raphael.deimel@tu-berlin.de)}
\address[Fourth]{Technische Universität Berlin, Control Systems Group (e-mail: seel@control.tu-berlin.de)}
\begin{abstract}
In motion tracking of connected multi-body systems Inertial Measurement Units (IMUs) are used in a wide variety of applications, since they provide a low-cost easy-to-use method for orientation estimation. However, in indoor environments or near ferromagnetic material the magnetic field is inhomogeneous which limits the accuracy of tracking algorithms using magnetometers. Methods that use only accelerometers and gyroscopes on the other hand yield no information on the absolute heading of the tracked object. For objects connected by rotational joints with range of motion constraints we propose a method that provides a magnetometer-free, long-term stable relative orientation estimate based on a non-linear, window-based cost function. The method can be used for real-time estimation as well as post-processing. It is validated experimentally with a mechanical joint and compared to other methods that are used in motion tracking. It is shown that for the used test object, the proposed methods yields the best results with a total angle error of less than $4\g$ for all experiments.
\end{abstract}
\end{frontmatter}
\section{Introduction}
The tracking of orientations and positions of objects in three-dimensional space is an integral part of various control applications. Many mechanical systems consist of multiple connected objects of which the individual orientations must be known. There exist numerous different approaches for orientation tracking reaching from optical systems, over mechanical solutions to inertial sensors. Inertial Measurement Units (IMUs) have the advantage of being small, low-cost, have a wide field of application and they require no direct interaction with the object of interest. Therefore they are used in a wide variety of robotic and biomedical applications \cite{Review_WearableSensing,Miller2004mocap,Fong2010IMU}. IMUs usually consist of 3D accelerometers, 3D gyroscopes and 3D magnetometers. Sensor fusion of these measurements yields the orientation of the sensor with respect to a fixed inertial frame. This common inertial frame is necessary to determine the relationships between the individual bodies of a kinematic chain. However, 9D sensor fusion only yields accurate orientation estimates if the magnetic field is homogeneous. In indoor environments and near ferromagnetic material the magnetic field is known to be highly disturbed and inhomogeneous \cite{subbu2013locateme,devries2009magnetic,shu2015magicol,grand20123axis}. This is crucial for most robotic and biomedical applications and makes conventional 9D sensor fusion inapplicable \cite{salchow2019tangible}. Without magnetometers, the heading component of the orientation is unknown. Without heading information, the estimated orientations of connected bodies cannot be used to determine relative orientations, joint angles or positional relationships. One approach by \cite{salchow2019tangible} is to omit the magnetometer readings and determine the heading by predefined initial poses. It is shown that the results are better than conventional 9D sensor fusion. However, due to drift this method only produces good results for short-term experiments ($<30\s$) and is not long-term stable. Therefore, current research aims to exploit the kinematic relationships between the connected bodies to obtain heading information without using magnetometers or predefined initial poses.
For joints with one degree of freedom, there exist methods to calculate the joint angles analytically without the use of magnetometers \cite{cooper2009inertial,seel2014imubased}. Recently a method was published which exploits the kinematic relationships with an orientation-based constraint \cite{laidig2017exploiting} to estimate the relative heading of two bodies connected by a one-dimensional joint. For joints with two-degrees of freedom, we recently published an orientation-based constraint which is independent from excitation and raw measurement data \cite{Laidig2019MagnetometerfreeRI}. % \item there are also methods that rely on the angular velocity \cite{laidig2017automatic} or on initial sensor-to-segment calibration protocols \cite{luinge2007ambulatory}, \cite{kortier2014assessment}
One drawback of those methods is that they require specific joint kinematics as well as known joint axes and sensor-to-segment orientations. If these requirements are met, these methods can produce accurate results. The methods for one and two-dimensional joints exploit the limited degrees of freedom to formulate kinematic constraints.

For three-dimensional joints, \cite{kok2014optimizationbased} and \cite{taetz2016selfcalibrating} exploit positional relationships and use those to estimate the true relative orientations of connected bodies. However, they still rely on magnetometers for the initialisation. In \cite{wenk2015posture} an approach is published which uses an EKF to estimate the orientation of two bodies connected by a ball joint. It is based on the knowledge of the relative position of the IMUs with respect to the joint center. These approaches for three-dimensional joints have in common that they rely on positional constraints and relationships or rely on raw measurement data for heading estimation.

In this paper, we propose a magnetometer-free method for relative motion tracking of kinematic chains with arbitrary joints that have limited range of motion (ROM). The method will be applicable to any rotational (1D, 2D and 3D) joints with ROM constraints, is only based on orientations and does not need known positional relationships or known joint axes. It further does not need any knowledge on initial orientation and yields a long-term stable estimation even under realistic conditions. The proposed method shows remarkable similarities to a Moving Horizon Estimation (MHE) and can be formulated as such. However, we derive and formulate the method in a more intuitive way.
\section{Kinematics and generic joint model}
\begin{figure}[!htp]
        \tikzsetnextfilename{kinematic_chain}
        \centering
        {
            \input{./img/kinematics/kinematic_chain.tex}
        }
          \caption[]{Model of a kinematic chain consisting of bodies $\seg{}$ connected by joints $\joint{}{}$}\label{fig_kinchain}
\end{figure}
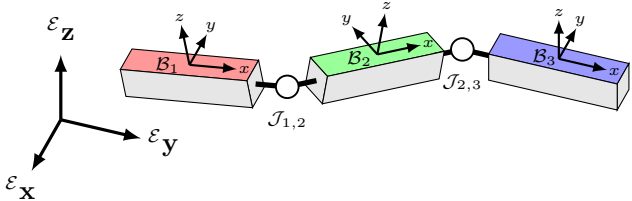
Consider a system of $N$ rigid bodies connected by $N-1$ joints. Rigid bodies are denoted $\seg{i}, \, i \in [1,N]$. The joint connecting the bodies $\seg{i}$ and $\seg{i+1}$ is denoted as $\joint{i}{i+1}$. Joints only connect two adjacent bodies to form a kinematic pair. An example for this system is shown in \figref{fig_kinchain}. We only consider rotational joints which allow for relative rotations of the two connected bodies with respect to each other. The orientation of each segment with respect to a reference frame is expressed by a quaternion. The global reference frame common for all bodies is denoted by $\earth{}$. The orientation of $\seg{i}$ with respect to $\earth{}$ is then denoted as $\quatsegearth{i}{}$.
The relative orientation $\quatsegseg{i+1}{i}$ of two adjacent bodies $\seg{i}$ and $\seg{i+1}$ describes the orientation of $\seg{i+1}$ with respect to the coordinate system of $\seg{i}$ and can be determined from the orientations of both bodies in the common reference frame:
  \begin{equation}\label{eq_relori}
  \quatsegseg{i+1}{i} = \quatsegearth{i}{}\quatinv \quatmult \quatsegearth{i+1}{}.
\end{equation}
The relative orientation of the two bodies is caused by the rotation around the joint connecting the two bodies. Depending on the mechanical model of a joint, it can either allow free relative rotation or the space of possible relative orientations is restricted. These restrictions can either be due to limited degrees of freedom (one- and two-dimensional joints), due to limitations of the range of motion or both. If the joint $\joint{i+1}{i}$ has restrictions of degrees of freedom or range of motion, the set of possible relative orientations $\quatsegseg{i+1}{i}$ for the joint is limited to a subset of all possible orientations $\mathbb{H}$. This subset is denoted as $\set{P}$ with
   \begin{equation}
    \set{P} \subseteq \mathbb{H}\,.
   \end{equation}
To describe the relative orientation of the two connected bodies, we propose a generic joint model based on the concept of \textit{joint axes}. We model the rotation $\quatsegseg{i+1}{i}$ from $\seg{i}$ to $\seg{i+1}$ as consecutive rotations around the joint axes $\jaxis{p} \in \mathbb{R}^3$, $\norm{\jaxis{p}} = 1$, by the \textit{joint angles} $\jangle{p} \in \mathbb{R}$, $\p \in [1\dots3]$. The restriction of the relative orientation is modeled as a limitation of the range of motion of one or more joint angles with
   \begin{equation}
     \jangle{p} \in \{\jangle{} \in \mathbb{R} | \varphi_{\p,\mathrm{min}} \leq \jangle{} \leq \varphi_{\p,\mathrm{max}} \} \,.
   \end{equation}
The relative orientation is then modeled as consecutive rotations around the joint axes
\small
\begin{equation}\label{eq_model}
  \quatsegseg{i+1}{i} = \quataa{\jangle{1}}{\jaxis{1}} \quatmult \quataa{\jangle{2}}{\jaxis{2}} \quatmult \quataa{\jangle{3}}{\jaxis{3}} = \prod_{p=1}^{3} \quataa{\jangle{p}}{\jaxis{p}} \,.
\end{equation} \normalsize
The operator \quataa{\alpha}{\jaxis{}} returns the quaternion describing the rotation of $\alpha$ around the axis $\jaxis{}$. The operator $\quatmult$ denotes quaternion multiplication.

\begin{figure}[h]
    \scriptsize{
          \centering
          \hspace{1cm}
          \subfigure[Distinct joint axes for a three-dimensional joint]{\def\svgwidth{.18\textwidth} 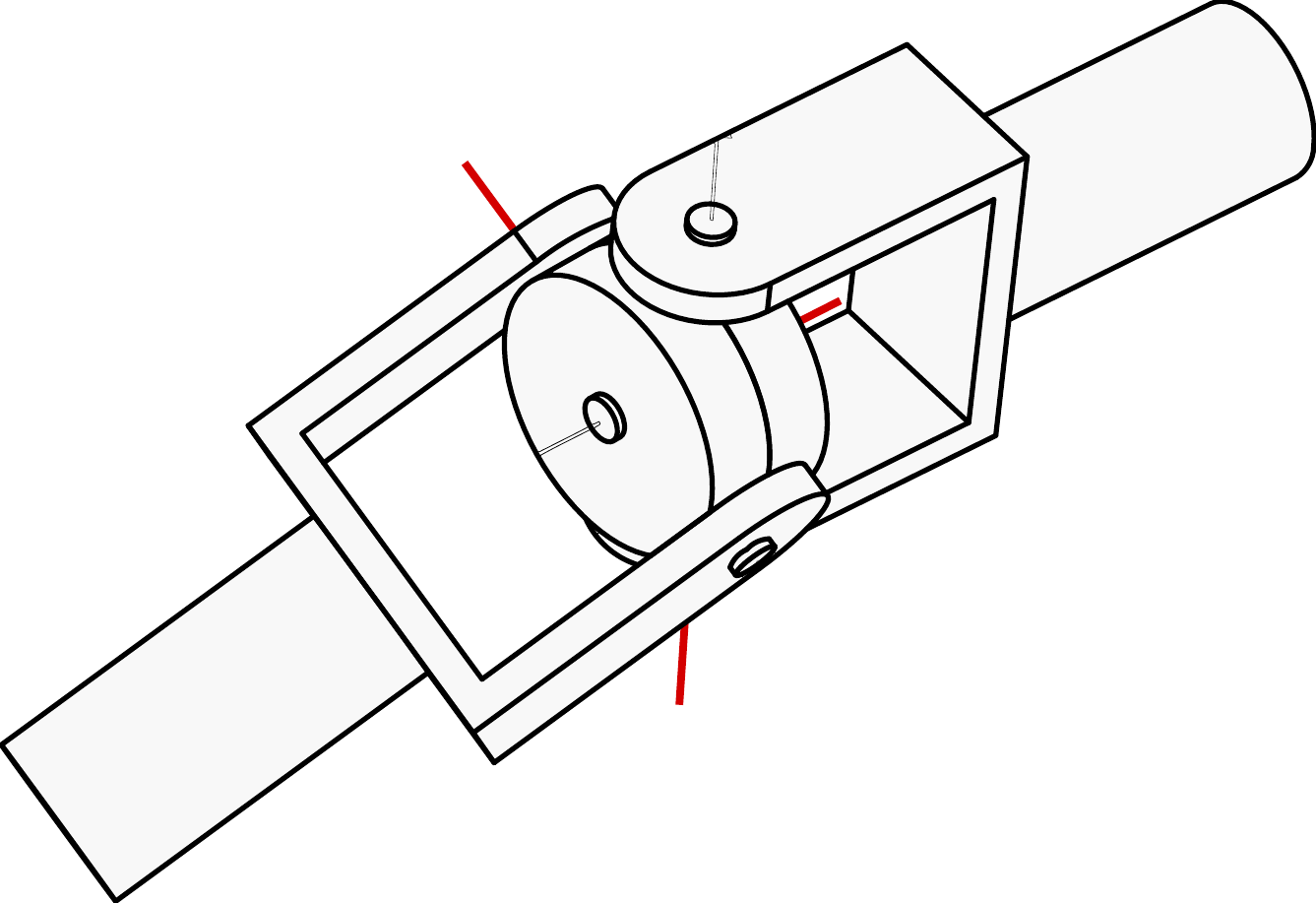}
          \hspace{0.2cm}
          \subfigure[Ball joint]{\def\svgwidth{.15\textwidth} 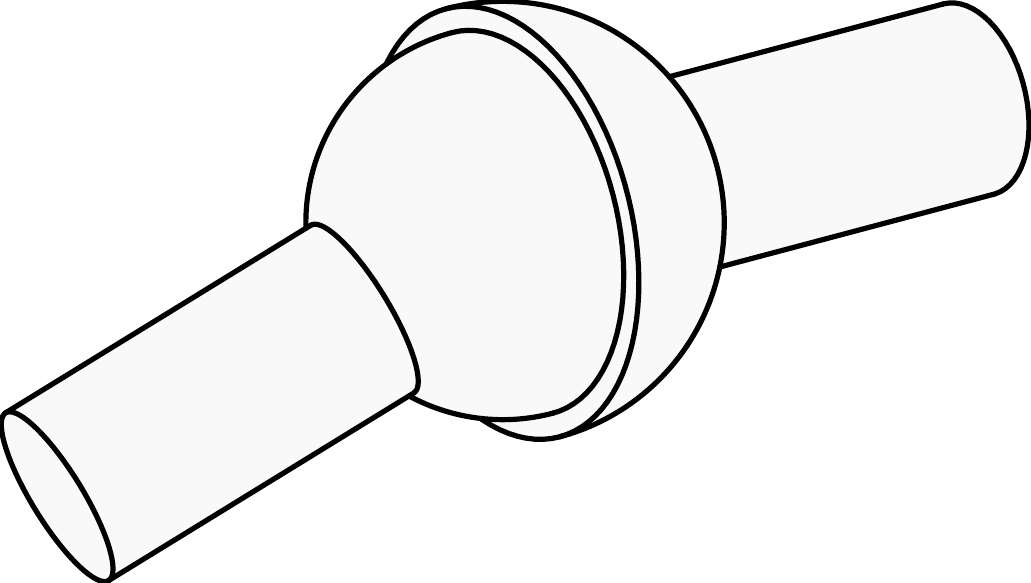}
          \hfill
          }\caption[]{Two joints with $\dof{\joint{}{}} = 3$}\label{fig_joint_configurations}
\end{figure}

This model can describe most types of rotational joints. For one- and two-dimensional joints, the predefined joint axes are chosen as $\jaxis{p}$. For the remaining degrees of freedom, linearly independent axes are chosen with a fixed value for the corresponding joint angles, i.e. $\varphi_{\p,\mathrm{min}} = \varphi_{\p,\mathrm{max}}$. This limits the relative orientation to rotations around the predefined joint axes of those joints. For three-dimensional joints, the joint axes $\jaxis{p}$ are chosen either as the predefined axes if the joint is a serial composition of hinge joints (see \figref{fig_joint_configurations} left) or can be chosen freely for joints with no distinct axes (see \figref{fig_joint_configurations} right). The set which contains all orientations described by the model \eqref{eq_model} is denoted as $\set[M]{P}$ and is defined as
{\small
\begin{equation}\label{eq_model_set}
    {
    \set[M]{P}\!:= \!\left\{\!\quat{}{} \!\in\mathbb{H} \middle| \quat{}{} = \prod_{\p=1}^{3} \quataa{\jangle{p}}{\jaxis{p}},\jangle{p}\!\in\! [\varphi_{\p,\mathrm{min}},\varphi_{\p,\mathrm{max}}]\!\right\}}.
\end{equation}}
The range of each joint angle follows from the mechanical model of the joint or from identification experiments.

Without loss of generality we will focus on the description of joints with three degrees of freedom since they describe the most general case. In \figref{fig_joint_configurations} two examples for joints with three degrees of freedom are shown.
For the left example (a), the joint axes $\jaxis{p}$ and the limits for the joint angles $\jangle{p}$ can be directly be extracted from the mechanical model. If we apply the model from \eqref{eq_model} to this, we can see that $\set[M]{P}$ perfectly describes the set of possible relative orientations, therefore $\set{P} = \set[M]{P}$. For the three-dimensional ball joint (b) any triplet of linearly independent joint axes can be chosen for the joint model. If we choose three axes $\vec{a},\vec{b}$ and $\vec{c}$ that form an orthonormal base of a right-handed coordinate system, we describe the joint by using Euler angles. This is a common approach and is simpler than choosing non-orthogonal axes. The corresponding Euler angles are denoted as $\alpha,\beta,\gamma \in \mathbb{R}$ and since the joint has range of motion constraints
\begin{equation}\label{eq_euler_limitations}
  \alpha \in [\alpha_{\min},\alpha_{\max}],\beta \in [\beta_{\min},\beta_{\max}],\gamma \in [\gamma_{\min},\gamma_{\max}] \,.
\end{equation}
If we apply the joint model to this, with $\vec{a}$, $\vec{b}$ and $\vec{c}$ being the joint axes $\jaxis{p}$ and $\alpha,\beta,\gamma$ the joint angles $\jangle{p}$, the set described by \eqref{eq_model_set} is a conservative approximation of $\set{P}$ since the joint model assumes fixed ranges for all joint angles. This creates a cuboid subspace in the space created by the three joint angles. For the ball joint however, not the complete subspace can be reached. Therefore, for this joint (and all joints in general)
\begin{equation}
  \set{P} \subseteq \set[M]{P} \subseteq \mathbb{H} \,.
\end{equation}
Describing each joint with a set of orthogonal axes and Euler angles can be easily applied to most types of rotational joints which have no distinct joint axes.

In the following derivation we only focus on a single kinematic pair of a kinematic chain with $N$ bodies. The proposed method can then be applied to each joint which fulfills the assumptions to estimate the orientations of the complete kinematic chain. Each joint estimation is independent from the other joints in the chain.
\section{Inertial state estimation}
    \begin{figure}[h]
      \centering
      {\small
      \def\svgwidth{.4\textwidth}
      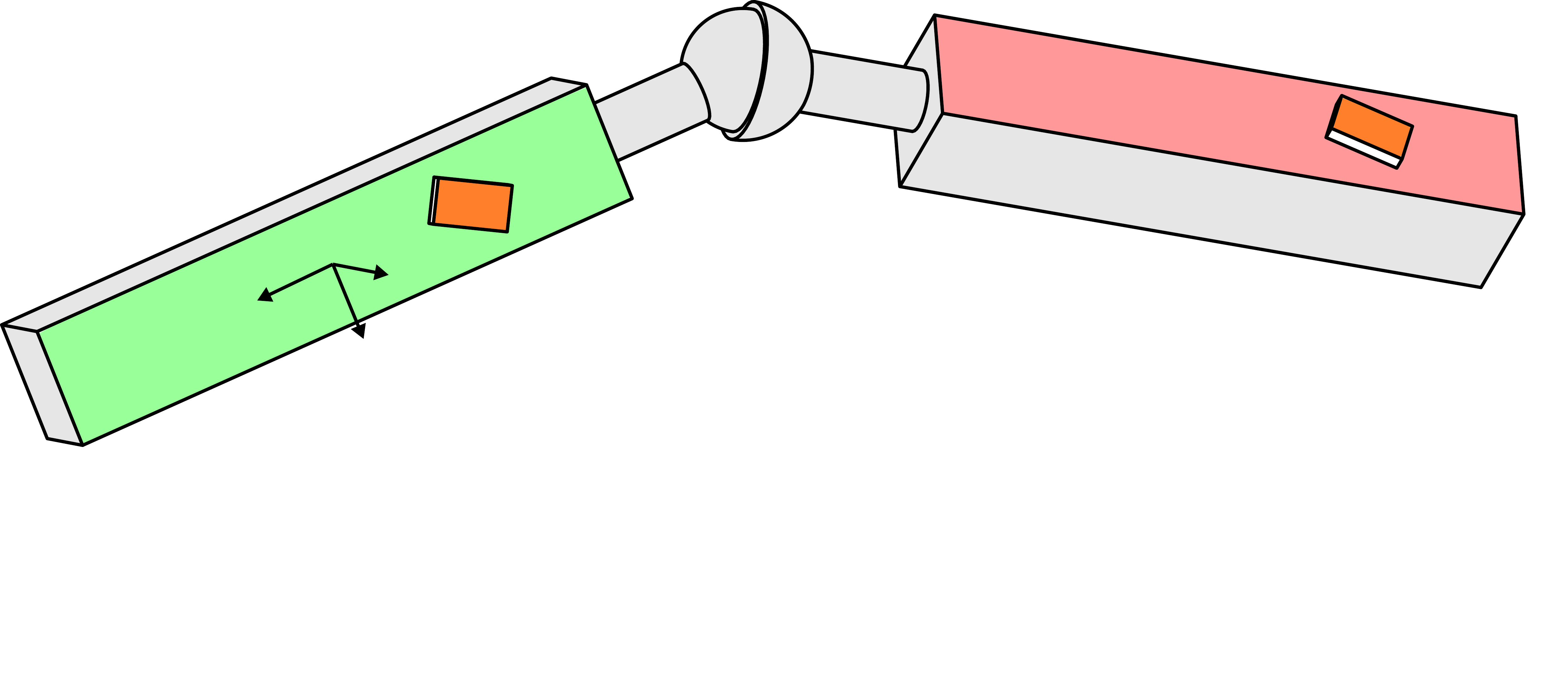
      }
      \caption[]{Kinematic model of two adjacent bodies $\seg{1}$ and $\seg{2}$ connected by the joint $\joint{1}{2}$ and the reference frame $\earth{1}$}\label{fig_initial_situation}
      \end{figure}
\begin{figure}[h]
        \centering
        {
            \input{./img/kinematics/kinematic_chain_delta.tex}
        }
          \caption{Difference between the two reference frames \earth{1} and \earth{2}}\label{fig_delta}
\end{figure}
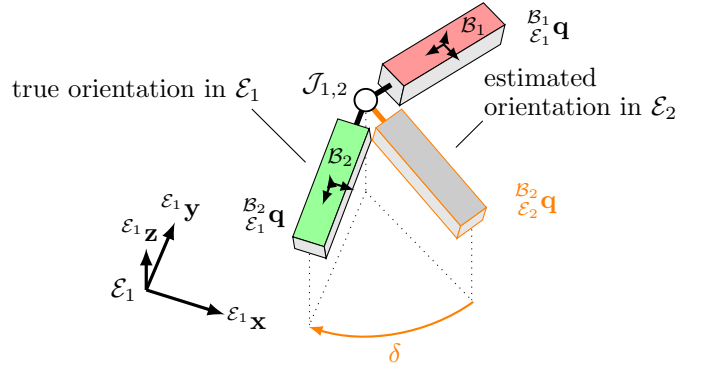

Consider two bodies $\seg{1}$ and $\seg{2}$ connected by the three-dimensional joint $\joint{1}{2}$ with range of motion constraints. To estimate the orientations of the bodies with respect to a reference frame, two inertial sensors $\imu{1}$ and $\imu{2}$ are placed on the bodies in a known and fixed orientation (see \figref{fig_initial_situation}). The sensors measure the angular rates $\gyr{1}{}(t)$ and $\gyr{2}{}(t)$ as well as the accelerations $\acc{1}{}(t)$ and $\acc{2}{}(t)$ in local coordinates of the sensors with a fixed sampling time $\TSample$. For each segment, 6D quaternion-based sensor fusion is performed for example according to the algorithm described in \cite{seel2017eliminating}, to obtain the orientation of each segment with respect to a reference frame, i.e. $\quatsegearth{1}{1}(t)$ and $\quatsegearth{2}{2}(t)$. However, due to the fact that no magnetometer is used and the resulting lack of heading information, the absolute heading of each segment at the beginning of the measurement is arbitrary and only dependent on the initial conditions of the sensor fusion algorithm. This can be modeled as if the orientations of the bodies are estimated in different reference frames $\earth{1}$ and $\earth{2}$. Since only the heading of each segment is unknown, the difference between the two reference frames $\earth{1}$ and $\earth{2}$ can be described by a rotation around the global vertical axis (see \figref{fig_delta}). The angle of this rotation is denoted as $\he$ and is called \textit{heading offset} \cite{laidig2017exploiting}. This rotation is described by the quaternion
\begin{equation}
  \quatearthearth{2}{1}(\he) = \quatcomp{\quatcos{\he}}{0}{0}{\quatsin{\he}}\,.
\end{equation}
With the two estimated orientations $\quatsegearth{1}{1}(t)$ and $\quatsegearth{2}{2}(t)$ and a known value of the heading offset $\he$ we can determine the relative orientation of the two bodies by
\begin{equation}
  \quatsegseg{2}{1}(t) = \quatsegearth{1}{1}(t)\quatinv \quatmult \quatearthearth{2}{1}(\he) \quatmult \quatsegearth{2}{2}(t) \,.
\end{equation}
Due to bias in the gyroscope measurements and the resulting drift around the global vertical axis, the angle $\he$ is not constant and can be approximated as
\begin{equation}\label{eq_delta_time}
          \delta(t) \approx \underbrace{\partialfrac{\delta}{t} \cdot t}_{\text{heading drift}} + \underbrace{\vphantom{\frac{a}{b}}\delta(t=0)\,.}_{\text{initial offset $\he_0$}}
\end{equation}
$\partialfrac{\he}{t}$ is a scalar which is slowly changing due to bias instability and non-uniformity of the bias of each axis. For small time windows the heading offset $\he(t)$ can be approximated constant, since the drift is typically small with magnitudes of $\partialfrac{\he}{t} < 0.5\,\frac{^\circ}{\mathrm{s}}$. Note that we do not make any other assumptions on the course of $\he(t)$.
\section{Constraint}
Consider two bodies $\seg{1}$ and $\seg{2}$ connected by the joint $\joint{}{}$ with the most general case of three degrees of freedom. The joint has mechanical restrictions which limits the set of possible relative orientations to $\set{P} \subset \mathbb{H}$. The kinematics of the joint are described by an Euler angle convention with the orthonormal axes $\vec{a}$, $\vec{b}$ and $\vec{c}$ and the corresponding angles $\alpha$, $\beta$, $\gamma$, which are restricted according to \eqref{eq_euler_limitations}. The set of relative orientations described by this model is $\set[M]{P}$ with $\set{P} \subseteq \set[M]{P}$. At any given point in time $t_k$, the orientations of the two segments estimated by the 6D sensor fusion are $\quatsegearth{1}{1}(t_k)$ and $\quatsegearth{2}{2}(t_k)$. The heading offset at that time instant is $\he(t_k)$. Let $\heest$ be an estimate of $\he(t_k)$, then an estimate $\quatsegsegest{2}{1}$ of the relative orientation can be obtained by
\begin{equation}
  \quatsegsegest{2}{1}(t_k,\heest) := \quatsegearth{1}{1}(t_k)\quatinv \quatmult \quatearthearthest{2}{1}(\heest) \quatmult \quatsegearth{2}{2}(t_k)\,.
\end{equation}
Let $E_{\mathrm{abc}}: \mathbb{H} \mapsto \mathbb{R}^3$ be an operator that maps any given quaternion $\quat{}{}$ to a triplet of Euler angles $(\alpha,\beta,\gamma)$ according to the intrinsic Euler angles convention $\euleri{a}{b}{c}$ \cite{Diebel2006RepresentingA}.
To check whether $\quatsegsegest{2}{1}(t_k,\heest)$ is an element of $\set[M]{P}$ at a given time instant $t_k$ and a given estimate $\heest$, it has to be decomposed into the Euler angles $\hat{\alpha}$, $\hat{\beta}$ and $\hat{\gamma}$ corresponding to the Euler angle convention $\euleri{a}{b}{c}$
\begin{equation}
  (\hat{\alpha},\hat{\beta},\hat{\gamma}) = E_{\mathrm{abc}}\left(\quatsegsegest{2}{1}(t_k,\heest)\right)\,.
\end{equation}
If $\hat{\alpha}$, $\hat{\beta}$ and $\hat{\gamma}$ are within the specified ranges according to \eqref{eq_euler_limitations}, then it is a valid relative orientation $\quatsegsegest{2}{1}(t_k,\heest) \in \set[M]{P}$ according to the model.
We define a function $e_k(\heest): \mathbb{R} \mapsto \mathbb{N}$ that assigns a binary value $0$ or $1$ to an estimate $\heest$ at a given time instant $t_k$ with
\begin{equation}\label{eq_cost_arbitrary}
      e_k(\heest) := \begin{cases}
                    0, & \mbox{if } \quatsegsegest{2}{1}(t_k,\heest) \in \set[M]{P}  \\
                    1, & \mbox{otherwise}.
                  \end{cases}
\end{equation}
\section{Estimation principle}
The basic idea of the estimation of $\he(t)$ is, that at each time instant $t_k$, only a subset of possible values of $\heest$ produces a relative orientation $\quatsegsegest{2}{1}(t_k,\heest)$ according to \eqref{eq_relori}, that lies within the set of possible relative orientations $\set[M]{P}$. With the assumption, that $\he(t)$ can be approximated constant over a small time window and the assumption of sufficiently rich motion, only one value of $\heest$ produces valid relative orientations for all time instants of that time window, i.e. does not violate the constraint. We therefore use a window-based approach to find a good estimate of $\he(t)$ for a given time interval $\TWin$.

At regular time intervals $\TEst \geq \TSample$, an estimation of $\he(t)$ is performed at the time instants $\windowvar{t}$, which are denoted with an index $w \in \mathbb{N}^+$ and are defined as
\begin{equation}
  \windowvar{t} := w \TEst\,.
\end{equation}
Each time window consists of $\windowsize$ samples taken at the sampling interval $\TSample$ at the sampling instants $\samplevar{t}$. The number of samples within a window is denoted as $\windowsize$. The time instants corresponding to the time window $w$ are given as
\begin{equation}\label{eq_sample_window_time_back}
  \samplevar{t} = \windowvar{t} + (m-\windowsize)\TSample, \,\,\, m\in[1\dots\windowsize]\,.
\end{equation}
We choose $\TEst < \TWin$ to create overlapping time windows. The time window definition only uses samples before the current estimation time instant, making the method real-time capable.

We assume that for sufficiently small time windows the heading offset $\he(t)$ can be approximated as constant. The estimate for the heading offset during a given time window $w$ is denoted by $\windowvar{\heest}$. Following this assumption and the model of $\he(t)$ from \eqref{eq_delta_time}, we assume that from one time window to the next one the value $\windowvar{\heest}$ does not change rapidly and is close to the previous value $\windowvar[w-1]{\heest}$. We formulate a cost function $c_w(\heest): \mathbb{R} \mapsto \mathbb{R}$ that for a given time window $w$ assigns a cost to a value of $\heest$
\begin{equation}\label{eq_cost}
  c_w(\heest) :=  \frac{\windowvar{N}}{\pi} \lvert(\heest - \windowvar[w-1]{\heest})\rvert + \sum_{m = 1}^{\windowvar{N}} \samplevar{e}(\heest) \,,
\end{equation}
with $k = w\TEst + (m-\windowvar{N})$.

The first term penalizes the distance to the previous estimate $\windowvar[w-1]{\heest}$. This ensures that the new estimate $\windowvar{\heest}$ is close to the previous estimate. The scaling factor $\frac{\windowvar{N}}{\pi}$ scales the cost that a distance of $\pi$ is equal to the maximum cost of the second term. We only penalize the distance and do not make assumptions on the direction of change. The second term is a measure of how well the estimate $\heest$ fulfills the kinematic constraint. The more valid relative orientations it produces over the course of the time window, the smaller the second term gets. \eqref{eq_cost} is then used to find an estimate $\windowvar{\heest}$ for a time window that minimizes the cost over that window by the help of any optimization method
\begin{equation}\label{eq_optimization}
  \windowvar{\heest} := \argmin{\heest} \,\, \windowvar{c}(\heest)\,.
\end{equation}
Note that the proposed method can be interpreted as a moving-horizon estimation approach for a dynamical system with one state $\delta$, which has an uncertain but small time derivative, and a highly non-linear output \eqref{eq_cost_arbitrary}, which we know to be zero up to small inaccuracies. The cost function \eqref{eq_cost} combines penalizes outputs different from this virtual zero measurements as well as state values that disagree with the uncertain dynamics.

For any time instant $t_k$ the value of the estimated heading offset $\heest(t_k)$ can be determined with
\begin{equation}
  \heest(t_k) = \heest_{\tilde{w}} \quad \text{with} \quad \tilde{w} = \left \lfloor{\frac{k}{\windowsize}}\right \rfloor \in \mathbb{N}^+ \,.\end{equation}

\begin{figure}[!htp]
    \tiny{
          \setlength{\figW}{2.3cm}
          \setlength{\figH}{1.5cm}
          \centering
          \hfill
          \subfigure{\input{./img/experiments/example_delta_3D_1.tex}}
          \subfigure{\input{./img/experiments/example_delta_3D_3.tex}}
          \subfigure{\input{./img/experiments/example_delta_3D_2.tex}}
          \hfill
          } \caption{Examples for cost functions $c_w(\heest)$ over $\heest \in [0,2\pi]$}\label{fig_examples_ew}
    \end{figure}
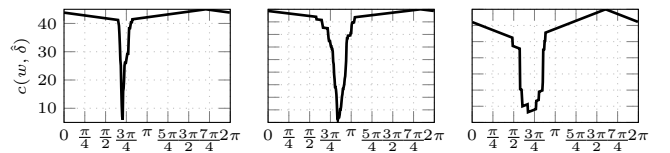

In \figref{fig_examples_ew} three examples for possible cost functions $c_w(\heest)$ over $\heest \in [0,2\pi]$ are shown. In the left graph, the constraint part of the cost function dominates. In the right graph, the distance cost to the previous estimate has more impact and ensures a non-diverging estimate even if the constraint minimum is less distinct.
With an estimate $\heest(t)$, both estimated orientations $\quatsegearth{1}{1}(t)$ and $\quatsegearth{2}{2}(t)$ can be transformed into a common reference frame. This allows us to calculate the relative orientation, joint angles and positional relationships between the connected bodies or to transform all body orientations of a kinematic chain into one common reference frame.
\section{Experimental validation}
The proposed method is validated experimentally with the use of an mechanical three-dimensional joint as a test object. As validation method an optical motion capture system is used to measure the relative orientation with very high precision. We investigate different motion patterns and speeds to evaluate the robustness of the method.
\subsection{Setup}
\begin{figure}[h]
        \centering
        \small{
            \def\svgwidth{.45\textwidth}
            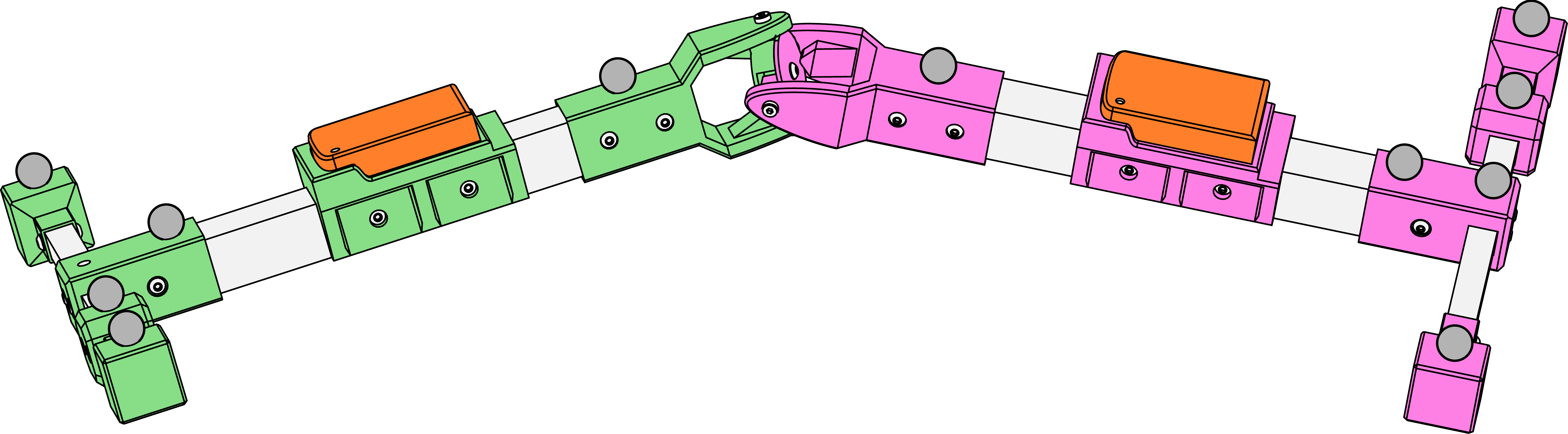
        }
        \caption{Mechanical model used for experimental evaluation. The two bodies are connected by a well-defined 3D joint. On each segment IR-reflective markers are attached for optical motion capture.}\label{fig_test_object}
\end{figure}

The test object is 3D printed (see \figref{fig_test_object}) with a well-defined three-dimensional joint connecting the two bodies. The range of motion of the joint angles are based on the range of motion of the Carpometacarpal joint of the thumb \cite{salchow2019tangible} of the author.
The joint is modeled by a \euleri{z}{x}{y} Euler angles convention with the angle ranges
    \begin{equation}
    \alpha \in [-20\g,20\g], \,\,\, \beta \in [-15\g,15\g], \,\,\, \gamma \in [-40\g,40\g] \,.
    \end{equation}
The joint axes $\jaxis{1}$, $\jaxis{2}$ and $\jaxis{3}$ are known and coincide with the axes of the $\euleri{z}{x}{y}$ Euler rotation sequence. On each segment, an IMU is placed in a known and constant orientation. The IMUs measure the angular velocity and the acceleration of each body at a rate of $75$ Hz and 6D sensor fusion is performed using the algorithm presented in \cite{seel2017eliminating}.
\subsection{Optical reference}
As a ground truth for the relative orientation of the two bodies an optical motion capture system (MoCap) is used. It consists of ten \textit{Optitrack Flex 13} cameras, attached in a rectangular pattern at the top of the measurement space. On the bodies 11 (5+6) IR-reflective markers are attached in a well-defined and known pattern (see \figref{fig_test_object}). The MoCap systems determines the 3D positions of the markers at a rate of $120$ Hz with a mean position error of less than $0.3$ mm. With the marker positions, each segment's orientation can be calculated at each sample time instant in a common reference frame and with \eqref{eq_relori} the true relative orientation of both bodies can be determined.
\subsection{Validation metrics}
Let $\quatsegsegest{2}{1}(t)$ be the estimated relative orientation of the two bodies determined by the algorithm described above. Let $\quatsegseg{2}{1}(t)$ be the true relative orientation determined by the MoCap system. Both timeseries are in a common timeframe.
We define the \textit{orientation error} $\epsilon(t)$ to be the error angle between the estimated and true relative orientation, i.e.
  \begin{equation}\label{eq_epsilon}
  \epsilon(t) := \left|\angle{\quatsegseg{2}{1}(t)\quatinv\quatmult\quatsegsegest{2}{1}(t)}\right|,
\end{equation}
with $A(\quat{}{})$ being the operator that extracts the angle of the corresponding quaternion $\quat{}{}$.
To quantify the method's ability to estimate the heading offset $\he(t)$ we introduce the error $\ehe(t)$, which is defined as the difference between the true and estimated value of $\he(t)$
\begin{equation}
  \ehe := |\he(t) - \heest(t)|\,.
\end{equation}
The error $\ehe(t)$ is the heading component of $\epsilon(t)$ after being transformed into the global reference frame. To quantify the overall performance over the course of an experiment, we use the metrics $\te$ and $\de$, which are the RMSE of their corresponding error metric.
\subsection{Conducted experiments}
Multiple short-term and long-term experiments have been conducted to test the initial estimation as well as the long-term stability. At the beginning of each experiment, the test object is resting in a random orientation on a table. It is then picked up and rotated and translated within the measurement space of the MoCap system. In all experiments the magnitude and frequency of excitation have been varied. A list of the experiments with their duration is given in \tabref{tab_experiments}.
\begin{table}[ht]
  \caption{Performed experiments}\label{tab_experiments}
  \centering
  \begin{tabular}{p{1.5cm}p{1.5cm}p{4cm}}
     \toprule
     \textbf{Experiment} & \textbf{Duration} & \textbf{Remark} \\ \hline \\[-6pt]
     $\mathrm{E}\_01$ & 64\s & random starting orientation \\
     $\mathrm{E}\_02$ & 62\s & random starting orientation\\
     $\mathrm{E}\_03$ & 64\s & random starting orientation\\
     $\mathrm{E}\_04$ & 301\s & fast movement\\
     $\mathrm{E}\_05$ & 212\s & slow movement\\
     $\mathrm{E}\_06$ & 305\s & mixed movement with pauses\\
     $\mathrm{E}\_07$ & 418\s & very long measurement with mixed movement\\
     \bottomrule
   \end{tabular}
\end{table}
For the estimation we use a window time $\TWin = 8\s$ and an estimation time $\TEst = 1\s$.
\subsection{Results}
\begin{figure}[h]
              {\scriptsize
              \setlength{\figW}{.18\textwidth}
              \setlength{\figH}{2cm}
              \centering
              \hfill
              \subfigure{\input{./img/experiments/log_plot_1.tex}}
              \subfigure{\input{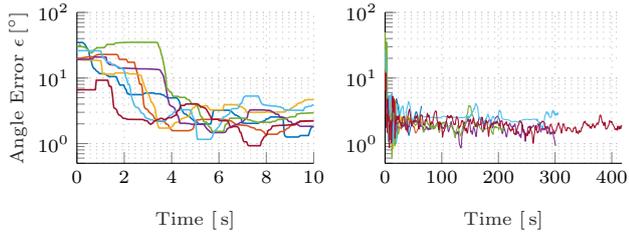}}
              \hfill
              }\caption{Error $\epsilon(t)$ for all experiments. Left: initial convergence during the first 10 seconds. Right: Error of the complete experiments.}\label{fig_initial_convergence}
\end{figure}

The method has to fulfill two objectives: converge towards the initial heading offset $\he_0$ at the beginning of the estimation and therefore the initial relative orientation as well as track the course of $\he(t)$ over an arbitrary length of the experiment for long-term stability. In \figref{fig_initial_convergence} the error $\epsilon(t)$ for all experiments is shown for the first 10 seconds as well as for the complete experiments.
Despite initial errors $>20\g$, in all cases the converges below $5\g$ within $5\s$ and drops below $4\g$ after $t>60\s$. The mean RMS for all experiments is $\te = 1.9\g$ with a heading tracking error of $\de = 0.8\g$. This shows that the method can estimate the true relative orientation very fast at the beginning of each estimation, which eradicates the need for predefined initial poses. The method is also able to track the relative orientation with a high accuracy even for long experiments and without external heading correction.

\begin{figure}[h]
        \centering
        \scriptsize{
            \setlength{\figW}{7cm}
            \setlength{\figH}{3cm}
            \input{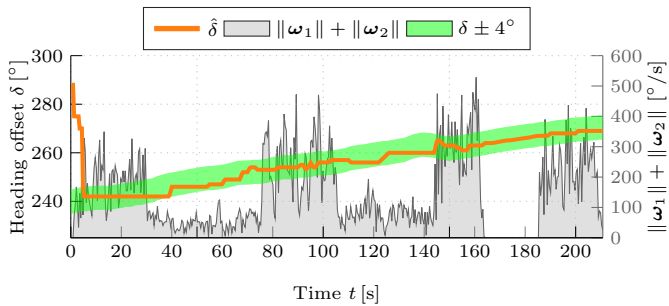}
        }
  \caption{Estimate $\heest(t)$ and reference value $\he(t)$ for experiment $\mathrm{E}\_05$  as well as norm of gyroscope measurements. }\label{fig_example_delta}
    \end{figure}

In \figref{fig_example_delta} the estimation of $\he(t)$ for one of the experiments is shown. The estimate follows the reference accurately with $\de = 0.9\g$. During the interval $t \in [164\s,184\s]$ the test object is lying down with no excitation. During that phase the first term of \eqref{eq_cost} ensures that the estimate does not diverge and stays close to the true value of $\he(t)$.

\begin{figure}[h]
        \centering
        \scriptsize{
            \setlength{\figW}{7cm}
            \setlength{\figH}{1.5cm}
            \input{./img/experiments/comparison.tex}
        }\caption{Errors $\epsilon(t)$ for the three evaluated methods for the experiment $\mathrm{E}\_05$.}\label{fig_comparison}
    \end{figure}
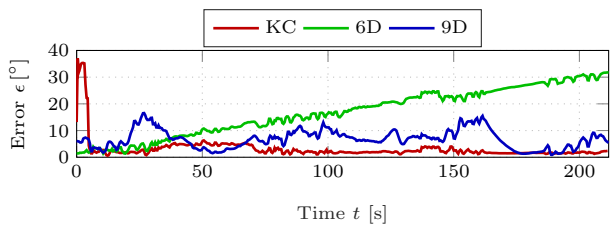

In \figref{fig_comparison} the error $\epsilon(t)$ is shown for the proposed method (KC) and, as a benchmark, 6D sensor fusion without constraints and only initial heading correction and bias compensation (6D) \cite{salchow2019tangible} as well as conventional 9D sensor fusion with magnetometer correction (9D). The proposed method performs best, with $\te = 2.2\g$. The 6D method produces adequate results at the beginning of the estimation but due to drift the relative orientation diverges, leading to a maximum error of $34\g$. Conventional 9D sensor fusion produces temporarily accurate results, but due to magnetic disturbances it has a RMS of $\te = 6.8\g$ with a maximum error of $18\g$.
The mean results for all experiments are given in \tabref{table_results}. Additionally the results for the conventional 6D and 9D sensor fusion methods are presented. The values in parentheses indicate the maximum errors $\max(\epsilon(t))$ and $\max(\ehe(t))$ of all experiments.
\begin{table}[ht]
\caption{Mean (Max) errors for all experiments}\label{table_results}
\centering
\begin{tabular}{p{0.6cm}p{2.0cm}p{2.0cm}p{1.5cm}}
  \toprule
  \textbf{Error} & \textbf{KC} & \textbf{6D} & \textbf{9D} \\  \midrule
  $\te$ & $\mathbf{2.1\g}$\,$(4\g)$ & $6.1\g$\,$(34\g)$ & $6.3\g$\,$(21\g)$\\
  $\de$ & $\mathbf{0.8\g}$\,$(2\g)$ & $5.4\g$\,$(32\g)$ & $5.9\g$\,$(20\g)$\\
  \bottomrule
\end{tabular}
\end{table}
The proposed method produces the best results with mean orientation errors of $2.1\g$ and maximum orientation errors of $4\g$. The objective of the method to track the heading offset can be accomplished with a mean error of less than $1\g$ for all experiments. The two conventional methods produce worse results with mean errors of approx. $6\g$ and maximum errors larger than $20\g$.
\section{Conclusion}
We proposed a new method for magnetometer-free inertial motion tracking for arbitrary joints with limited range of motion. The method exploits ROM constraints and the limited set of possible relative orientations to estimate the relative heading of two connected bodies using a window-based approach. To apply the method to different joint geometries we proposed a generic joint model for rotational joints which can sufficiently approximate joints with no distinct joint axes.

The method enables magnetometer-free real-time tracking in real-world indoor environments. We showed that the method yields an accurate long-term stable estimate for a case study derived from a real physiological joint. Compared to other methods \cite{salchow2019tangible} that also omit the magnetometer readings, the method does not rely on known initial poses and converges towards the true relative orientation within seconds. We showed for the evaluated test object that the methods yields better results than conventional 9D motion tracking and is long-term stable in contrast to 6D motion tracking methods.

However, compared to different orientation-based constraints that exploit the limited degrees of freedom of one- and two-dimensional joints \cite{laidig2017exploiting,Laidig2019MagnetometerfreeRI}, the method needs recurrent excitation. Furthermore, the method only works for joints where at least one of the joint angles has a moderate range of motion restriction. It has to be ensured to find the Euler angle convention which produces the smallest ranges for the joint angles.

Future work aims at overcoming the set limitations, i.e. investigating the sufficiency of rich motion, the magnitude of the angular restrictions as well as validating the method in less rigid, biological joints.
\begin{ack}
We gratefully acknowledge financial support for the project MTI-engAge (16SV7109) by BMBF.
\end{ack}
\bibliography{Paper_3D_IFAC_WC}
\end{document}

%% file: img/kinematics/kinematic_chain.tex
\tdplotsetmaincoords{50}{110}
\pgfdeclarelayer{back}
\pgfdeclarelayer{middle}
\pgfdeclarelayer{front}
\pgfsetlayers{back,main,middle,front}

\makeatletter
\pgfkeys{%
  /tikz/on layer/.code={
    \pgfonlayer{#1}\begingroup
    \aftergroup\endpgfonlayer
    \aftergroup\endgroup
  },
  /tikz/node on layer/.code={
    \gdef\node@@on@layer{%
      \setbox\tikz@tempbox=\hbox\bgroup\pgfonlayer{#1}\unhbox\tikz@tempbox\endpgfonlayer\egroup}
    \aftergroup\node@on@layer
  },
  /tikz/end node on layer/.code={
    \endpgfonlayer\endgroup\endgroup
  }
}

\def\node@on@layer{\aftergroup\node@@on@layer}

\makeatother
\begin{tikzpicture}[scale=2.25,tdplot_main_coords]
\tikzset{>=latex}
  % variables
  \def\rvec{.8}
  \def\thetavec{30}
  \def\phivec{60}

    \def\jointdist{0.4}
  % axes
  \coordinate (O) at (0,0,0);
  \draw[very thick,->] (0,0,0) -- (.5,0,0) node[distance from end=-0.3cm]{$\vecframe{x}{\earth{}}{}$};
  \draw[very thick,->] (0,0,0) -- (0,.5,0) node[distance from end=-0.3cm]{$\vecframe{y}{\earth{}}{}$};
  \draw[very thick,->] (0,0,0) -- (0,0,.5) node[distance from end=-0.3cm]{$\vecframe{z}{\earth{}}{}$};

\pgfmathsetmacro{\cubex}{0.8}
\pgfmathsetmacro{\cubey}{0.2}
\pgfmathsetmacro{\cubez}{0.2}

\coordinate (OC) at (0.5,1,1);

% BOX 1
\tdplotsetrotatedcoordsorigin{(OC)}
\tdplotsetrotatedcoords{90}{-10}{0}
\node[draw=none,shape=circle,fill, inner sep=1pt,tdplot_rotated_coords] at (OC){};
\draw[black,fill=red!40,opacity=1,tdplot_rotated_coords] ($(OC)+(\cubex/2,\cubey/2,\cubez/2)$) -- ++(-\cubex,0,0) -- ++(0,-\cubey,0) -- ++(\cubex,0,0) -- cycle;
\draw[black,fill=black!10,opacity=1,tdplot_rotated_coords] ($(OC)+(\cubex/2,\cubey/2,\cubez/2)$) -- ++(0,0,-\cubez) -- ++(0,-\cubey,0) -- ++(0,0,\cubez) -- cycle;
\draw[black,fill=black!10,opacity=.0,tdplot_rotated_coords] ($(OC)+(\cubex/2,\cubey/2,\cubez/2)$) -- ++(-\cubex,0,0) -- ++(0,0,-\cubez) -- ++(\cubex,0,0) -- cycle;
\draw[black,fill=black!10,opacity=1,tdplot_rotated_coords] ($($(OC)+(\cubex/2,\cubey/2,\cubez/2)$)-(\cubex,\cubey,\cubez)$) -- ++(\cubex,0,0) -- ++(0,0,\cubez) -- ++(-\cubex,0,0) -- cycle;
\draw[black,fill=black!20,opacity=.0,tdplot_rotated_coords] ($($(OC)+(\cubex/2,\cubey/2,\cubez/2)$)-(\cubex,\cubey,\cubez)$) -- ++(0,\cubey,0) -- ++(\cubex,0,0) -- ++(0,-\cubey,0) -- cycle;
\draw[black,fill=black!20,opacity=.0,tdplot_rotated_coords] ($($(OC)+(\cubex/2,\cubey/2,\cubez/2)$)-(\cubex,\cubey,\cubez)$) -- ++(0,\cubey,0) -- ++(0,0,\cubez) -- ++(0,-\cubey,0) -- cycle;

% BOX 1 - BOX 2

\tdplottransformrotmain{\cubex/2}{0}{0}
\coordinate (end_1) at ($(\tdplotresx, \tdplotresy, \tdplotresz)+(OC)$);
\tdplottransformrotmain{0.75*\cubex}{0}{0}
\coordinate (joint_1) at ($(\tdplotresx, \tdplotresy, \tdplotresz)+(OC)$);
\draw[line width=2pt] (end_1) -- (joint_1);
\node[circle,draw=black, line width=0.25mm, inner sep=3pt,fill=white,node on layer=front] at (joint_1){};

% coordinate system
\tdplottransformrotmain{0}{0}{\cubez/2}
\coordinate (cs_1) at ($(\tdplotresx, \tdplotresy, \tdplotresz)+(OC)$);
\tdplotsetrotatedcoordsorigin{(cs_1)}
\draw[thick,->,tdplot_rotated_coords] (cs_1) -- (0.3,0,0) node [distance from end=-0.1cm]{\tiny{$x$}};
\draw[thick,->,tdplot_rotated_coords] (cs_1) -- (0,0.3,0) node[distance from end=-0.1cm]{\tiny{$y$}};
\draw[thick,->,tdplot_rotated_coords] (cs_1) -- (0,0,0.3) node[distance from end=-0.1cm]{\tiny{$z$}};
\node[anchor=east] at (cs_1){\scriptsize{$\seg{1}$}};
\tdplotsetrotatedcoordsorigin{(OC)}
% ---------------------------------------------------------------------------------------------%
% BOX 2
\tdplotsetrotatedcoords{140}{10}{0} % ORIENTATION OF BOX 2
\tdplotsetrotatedcoordsorigin{(joint_1)}
\tdplottransformrotmain{0.25*\cubex}{0}{0}
\coordinate (start_2) at ($(\tdplotresx, \tdplotresy, \tdplotresz)+(joint_1)$);

\node[below= 0.2cm of joint_1] {\scriptsize{$\joint{1}{2}$}};

\draw[line width=2pt,on layer=middle] (joint_1) -- (start_2);
\tdplottransformrotmain{0.75*\cubex}{0}{0}
\coordinate (OC2) at ($(\tdplotresx, \tdplotresy, \tdplotresz)+(joint_1)$);
% \node[draw=none,shape=circle,fill, inner sep=1pt,color=red] at (start_2){};
% \node[draw=none,shape=circle,fill, inner sep=1pt,color=red] at (OC2){};

\tdplotsetrotatedcoordsorigin{(OC2)}

\draw[black,fill=green!40,opacity=1,tdplot_rotated_coords] ($(OC2)+(\cubex/2,\cubey/2,\cubez/2)$) -- ++(-\cubex,0,0) -- ++(0,-\cubey,0) -- ++(\cubex,0,0) -- cycle;
\draw[black,fill=black!20,opacity=.0,tdplot_rotated_coords] ($(OC2)+(\cubex/2,\cubey/2,\cubez/2)$) -- ++(0,0,-\cubez) -- ++(0,-\cubey,0) -- ++(0,0,\cubez) -- cycle;
\draw[black,fill=black!20,opacity=.0,tdplot_rotated_coords] ($(OC2)+(\cubex/2,\cubey/2,\cubez/2)$) -- ++(-\cubex,0,0) -- ++(0,0,-\cubez) -- ++(\cubex,0,0) -- cycle;
\draw[black,fill=black!10,opacity=1,tdplot_rotated_coords,on layer=front] ($($(OC2)+(\cubex/2,\cubey/2,\cubez/2)$)-(\cubex,\cubey,\cubez)$) -- ++(\cubex,0,0) -- ++(0,0,\cubez) -- ++(-\cubex,0,0) -- cycle;
\draw[black,fill=black!20,opacity=.0,tdplot_rotated_coords] ($($(OC2)+(\cubex/2,\cubey/2,\cubez/2)$)-(\cubex,\cubey,\cubez)$) -- ++(0,\cubey,0) -- ++(\cubex,0,0) -- ++(0,-\cubey,0) -- cycle;
\draw[black,fill=black!10,opacity=1,tdplot_rotated_coords,on layer=back] ($($(OC2)+(\cubex/2,\cubey/2,\cubez/2)$)-(\cubex,\cubey,\cubez)$) -- ++(0,\cubey,0) -- ++(0,0,\cubez) -- ++(0,-\cubey,0) -- cycle;

\tdplottransformrotmain{\cubex/2}{0}{0}
\coordinate (end_2) at ($(\tdplotresx, \tdplotresy, \tdplotresz)+(OC2)$);
% \node[draw=none,shape=circle,fill, inner sep=1pt,color=red] at (end_2){};
\tdplottransformrotmain{0.75*\cubex}{0}{0}
\coordinate (joint_2) at ($(\tdplotresx, \tdplotresy, \tdplotresz)+(OC2)$);
\draw[line width=2pt] (end_2) -- (joint_2);
\node[circle,draw=black, line width=0.25mm, inner sep=3pt,fill=white,node on layer=front] at (joint_2){};
\node[below= 0.2cm of joint_2] {\scriptsize{$\joint{2}{3}$}};
% coordinate system
\tdplottransformrotmain{0}{0}{\cubez/2}
\coordinate (cs_2) at ($(\tdplotresx, \tdplotresy, \tdplotresz)+(OC2)$);
\tdplotsetrotatedcoordsorigin{(cs_2)}
\draw[thick,->,tdplot_rotated_coords] (cs_2) -- (0.3,0,0) node [distance from end=-0.1cm]{\tiny{$x$}};
\draw[thick,->,tdplot_rotated_coords] (cs_2) -- (0,0.3,0) node[distance from end=-0.1cm]{\tiny{$y$}};
\draw[thick,->,tdplot_rotated_coords] (cs_2) -- (0,0,0.3) node[distance from end=-0.1cm]{\tiny{$z$}};
\node[below left=-0.17cm and -0.05cm] at (cs_2){\scriptsize{$\seg{2}$}};
\tdplotsetrotatedcoordsorigin{(OC2)}

% ---------------------------------------------------------------------------------------------%
% BOX 3
\tdplotsetrotatedcoords{90}{0}{0} % ORIENTATION OF BOX 3
\tdplotsetrotatedcoordsorigin{(joint_2)}
\tdplottransformrotmain{0.25*\cubex}{0}{0}
\coordinate (start_3) at ($(\tdplotresx, \tdplotresy, \tdplotresz)+(joint_2)$);

\draw[line width=2pt] (joint_2) -- (start_3);
\tdplottransformrotmain{0.75*\cubex}{0}{0}
\coordinate (OC3) at ($(\tdplotresx, \tdplotresy, \tdplotresz)+(joint_2)$);

\tdplotsetrotatedcoordsorigin{(OC2)}

\draw[black,fill=blue!40,opacity=1,tdplot_rotated_coords] ($(OC3)+(\cubex/2,\cubey/2,\cubez/2)$) -- ++(-\cubex,0,0) -- ++(0,-\cubey,0) -- ++(\cubex,0,0) -- cycle;
\draw[black,fill=black!10,opacity=1,tdplot_rotated_coords] ($(OC3)+(\cubex/2,\cubey/2,\cubez/2)$) -- ++(0,0,-\cubez) -- ++(0,-\cubey,0) -- ++(0,0,\cubez) -- cycle;
\draw[black,fill=black!20,opacity=.0,tdplot_rotated_coords] ($(OC3)+(\cubex/2,\cubey/2,\cubez/2)$) -- ++(-\cubex,0,0) -- ++(0,0,-\cubez) -- ++(\cubex,0,0) -- cycle;
\draw[black,fill=black!10,opacity=1,tdplot_rotated_coords] ($($(OC3)+(\cubex/2,\cubey/2,\cubez/2)$)-(\cubex,\cubey,\cubez)$) -- ++(\cubex,0,0) -- ++(0,0,\cubez) -- ++(-\cubex,0,0) -- cycle;
\draw[black,fill=black!20,opacity=.0,tdplot_rotated_coords] ($($(OC3)+(\cubex/2,\cubey/2,\cubez/2)$)-(\cubex,\cubey,\cubez)$) -- ++(0,\cubey,0) -- ++(\cubex,0,0) -- ++(0,-\cubey,0) -- cycle;
\draw[black,fill=black!20,opacity=.0,tdplot_rotated_coords] ($($(OC3)+(\cubex/2,\cubey/2,\cubez/2)$)-(\cubex,\cubey,\cubez)$) -- ++(0,\cubey,0) -- ++(0,0,\cubez) -- ++(0,-\cubey,0) -- cycle;

% coordinate system
\tdplottransformrotmain{0}{0}{\cubez/2}
\coordinate (cs_3) at ($(\tdplotresx, \tdplotresy, \tdplotresz)+(OC3)$);
\tdplotsetrotatedcoordsorigin{(cs_3)}
\draw[thick,->,tdplot_rotated_coords] (cs_3) -- (0.3,0,0) node [distance from end=-0.1cm]{\tiny{$x$}};
\draw[thick,->,tdplot_rotated_coords] (cs_3) -- (0,0.3,0) node[distance from end=-0.1cm]{\tiny{$y$}};
\draw[thick,->,tdplot_rotated_coords] (cs_3) -- (0,0,0.3) node[distance from end=-0.1cm]{\tiny{$z$}};
\node[below left=-0.25cm and -0.1cm] at (cs_3){\scriptsize{$\seg{3}$}};
\tdplotsetrotatedcoordsorigin{(OC3)}

% \node[draw=none,shape=circle,fill, inner sep=1pt,tdplot_rotated_coords] at (OC2){};

%   \draw[thick,->,tdplot_rotated_coords] (OC) -- (0.4,0,0) node [distance from end=-0.3cm]{$\vecframe{x}{B}{}$};
%   \draw[thick,->,tdplot_rotated_coords] (OC) -- (0,0.4,0) node[distance from end=-0.2cm]{$\vecframe{y}{B}{}$};
%   \draw[thick,->,tdplot_rotated_coords] (OC) -- (0,0,0.4) node[distance from end=-0.1cm]{$\vecframe{z}{B}{}$};
%   \draw[->,blue,very thick] (O)  -- (O2) node [midway, below right=-0.2cm] {$\vecframe{p}{B}{A}$};
%
% \node[draw=none,shape=circle,fill, inner sep=1pt,tdplot_rotated_coords] at (OC){};
% \node[above left = -0.1cm of OC] {$\seg{}$};
% \draw[->,black,thick] (O)  -- (OC) node [midway, below right=-0.2cm] {$\vecframe{p}{B}{A}$};

% \draw[dashed, color=red] (O) -- (Pxy);
\end{tikzpicture} 

%% file: img/kinematics/Joint3.pdf_tex
%% Accompanies image file 'Joint3.pdf' (pdf, eps, ps)
\begingroup%
  \makeatletter%
  \providecommand\color[2][]{%
    \errmessage{(Inkscape) Color is used for the text in Inkscape, but the package 'color.sty' is not loaded}%
    \renewcommand\color[2][]{}%
  }%
  \providecommand\transparent[1]{%
    \errmessage{(Inkscape) Transparency is used (non-zero) for the text in Inkscape, but the package 'transparent.sty' is not loaded}%
    \renewcommand\transparent[1]{}%
  }%
  \providecommand\rotatebox[2]{#2}%
  \newcommand*\fsize{\dimexpr\f@size pt\relax}%
  \newcommand*\lineheight[1]{\fontsize{\fsize}{#1\fsize}\selectfont}%
  \ifx\svgwidth\undefined%
    \setlength{\unitlength}{384.89479065bp}%
    \ifx\svgscale\undefined%
      \relax%
    \else%
      \setlength{\unitlength}{\unitlength * \real{\svgscale}}%
    \fi%
  \else%
    \setlength{\unitlength}{\svgwidth}%
  \fi%
  \global\let\svgwidth\undefined%
  \global\let\svgscale\undefined%
  \makeatother%
  \begin{picture}(1,0.68583845)%
    \lineheight{1}%
    \setlength\tabcolsep{0pt}%
    \put(0,0){\includegraphics[width=\unitlength,page=1]{Joint3.pdf}}%
    \put(0.54584491,0.6184371){\color[rgb]{0.83137255,0,0}\makebox(0,0)[lt]{\lineheight{1.25}\smash{\begin{tabular}[t]{l}$\ja{1}$\end{tabular}}}}%
    \put(0,0){\includegraphics[width=\unitlength,page=2]{Joint3.pdf}}%
    \put(0.63538248,0.17722758){\color[rgb]{0.83137255,0,0}\makebox(0,0)[lt]{\lineheight{1.25}\smash{\begin{tabular}[t]{l}$\ja{3}$\end{tabular}}}}%
    \put(0.36080839,0.32262452){\color[rgb]{0.83137255,0,0}\makebox(0,0)[rt]{\lineheight{1.25}\smash{\begin{tabular}[t]{r}$\ja{2}$\end{tabular}}}}%
  \end{picture}%
\endgroup%

%% file: img/kinematics/Joint3_Ball.pdf_tex
%% Accompanies image file 'Joint3_Ball.pdf' (pdf, eps, ps)
\begingroup%
  \makeatletter%
  \providecommand\color[2][]{%
    \errmessage{(Inkscape) Color is used for the text in Inkscape, but the package 'color.sty' is not loaded}%
    \renewcommand\color[2][]{}%
  }%
  \providecommand\transparent[1]{%
    \errmessage{(Inkscape) Transparency is used (non-zero) for the text in Inkscape, but the package 'transparent.sty' is not loaded}%
    \renewcommand\transparent[1]{}%
  }%
  \providecommand\rotatebox[2]{#2}%
  \newcommand*\fsize{\dimexpr\f@size pt\relax}%
  \newcommand*\lineheight[1]{\fontsize{\fsize}{#1\fsize}\selectfont}%
  \ifx\svgwidth\undefined%
    \setlength{\unitlength}{296.70687103bp}%
    \ifx\svgscale\undefined%
      \relax%
    \else%
      \setlength{\unitlength}{\unitlength * \real{\svgscale}}%
    \fi%
  \else%
    \setlength{\unitlength}{\svgwidth}%
  \fi%
  \global\let\svgwidth\undefined%
  \global\let\svgscale\undefined%
  \makeatother%
  \begin{picture}(1,0.56576351)%
    \lineheight{1}%
    \setlength\tabcolsep{0pt}%
    \put(0,0){\includegraphics[width=\unitlength,page=1]{Joint3_Ball.pdf}}%
  \end{picture}%
\endgroup%

%% file: img/kinematics/Model_3D.pdf_tex
%% Accompanies image file 'Model_3D.pdf' (pdf, eps, ps)
\begingroup%
  \makeatletter%
  \providecommand\color[2][]{%
    \errmessage{(Inkscape) Color is used for the text in Inkscape, but the package 'color.sty' is not loaded}%
    \renewcommand\color[2][]{}%
  }%
  \providecommand\transparent[1]{%
    \errmessage{(Inkscape) Transparency is used (non-zero) for the text in Inkscape, but the package 'transparent.sty' is not loaded}%
    \renewcommand\transparent[1]{}%
  }%
  \providecommand\rotatebox[2]{#2}%
  \newcommand*\fsize{\dimexpr\f@size pt\relax}%
  \newcommand*\lineheight[1]{\fontsize{\fsize}{#1\fsize}\selectfont}%
  \ifx\svgwidth\undefined%
    \setlength{\unitlength}{1432.16574097bp}%
    \ifx\svgscale\undefined%
      \relax%
    \else%
      \setlength{\unitlength}{\unitlength * \real{\svgscale}}%
    \fi%
  \else%
    \setlength{\unitlength}{\svgwidth}%
  \fi%
  \global\let\svgwidth\undefined%
  \global\let\svgscale\undefined%
  \makeatother%
  \begin{picture}(1,0.43019701)%
    \lineheight{1}%
    \setlength\tabcolsep{0pt}%
    \put(0,0){\includegraphics[width=\unitlength,page=1]{Model_3D.pdf}}%
    \put(0.15927759,0.22673046){\color[rgb]{0,0,0}\makebox(0,0)[rt]{\lineheight{1.25}\smash{\begin{tabular}[t]{r}$\sz{x}$\end{tabular}}}}%
    \put(0.23282805,0.19389375){\color[rgb]{0,0,0}\makebox(0,0)[t]{\lineheight{1.25}\smash{\begin{tabular}[t]{c}$\sz{y}$\end{tabular}}}}%
    \put(0.25231375,0.24731932){\color[rgb]{0,0,0}\makebox(0,0)[lt]{\lineheight{1.25}\smash{\begin{tabular}[t]{l}$\sz{z}$\end{tabular}}}}%
    \put(0.21504823,0.27348228){\color[rgb]{0,0,0}\makebox(0,0)[lt]{\lineheight{1.25}\smash{\begin{tabular}[t]{l}${\seg{2}}$\end{tabular}}}}%
    \put(0.3360523,0.30677128){\color[rgb]{0,0,0}\makebox(0,0)[lt]{\lineheight{1.25}\smash{\begin{tabular}[t]{l}${\imu{2}}$\end{tabular}}}}%
    \put(0,0){\includegraphics[width=\unitlength,page=2]{Model_3D.pdf}}%
    \put(0.71729451,0.36223213){\color[rgb]{0,0,0}\makebox(0,0)[rt]{\lineheight{1.25}\smash{\begin{tabular}[t]{r}$\sz{x}$\end{tabular}}}}%
    \put(0.78547802,0.29181341){\color[rgb]{0,0,0}\makebox(0,0)[t]{\lineheight{1.25}\smash{\begin{tabular}[t]{c}$\sz{y}$\end{tabular}}}}%
    \put(0.81113652,0.40580239){\color[rgb]{0,0,0}\makebox(0,0)[t]{\lineheight{1.25}\smash{\begin{tabular}[t]{c}$\sz{z}$\end{tabular}}}}%
    \put(0.79981865,0.34390625){\color[rgb]{0,0,0}\makebox(0,0)[lt]{\lineheight{1.25}\smash{\begin{tabular}[t]{l}${\seg{1}}$\end{tabular}}}}%
    \put(0.90908224,0.32631412){\color[rgb]{0,0,0}\makebox(0,0)[lt]{\lineheight{1.25}\smash{\begin{tabular}[t]{l}${\imu{1}}$\end{tabular}}}}%
    \put(0,0){\includegraphics[width=\unitlength,page=3]{Model_3D.pdf}}%
    \put(0.85022148,0.19495434){\color[rgb]{0,0,0}\makebox(0,0)[t]{\lineheight{1.25}\smash{\begin{tabular}[t]{c}$\vecframe{z}{\earth{1}}{}$\end{tabular}}}}%
    \put(0.79033715,0.00187107){\color[rgb]{0,0,0}\makebox(0,0)[t]{\lineheight{1.25}\smash{\begin{tabular}[t]{c}$\vecframe{x}{\earth{1}}{}$\end{tabular}}}}%
    \put(1.00213426,0.04766502){\color[rgb]{0,0,0}\makebox(0,0)[t]{\lineheight{1.25}\smash{\begin{tabular}[t]{c}$\vecframe{y}{\earth{1}}{}$\end{tabular}}}}%
    \put(0.48756029,0.29210421){\color[rgb]{0,0,0}\makebox(0,0)[t]{\lineheight{1.25}\smash{\begin{tabular}[t]{c}${\joint{1}{2}}$\end{tabular}}}}%
  \end{picture}%
\endgroup%

%% file: img/kinematics/kinematic_chain_delta.tex
\tdplotsetmaincoords{30}{20}
\pgfdeclarelayer{back}
\pgfdeclarelayer{middle}
\pgfdeclarelayer{front}
\pgfsetlayers{back,main,middle,front}

\makeatletter
\pgfkeys{%
  /tikz/on layer/.code={
    \pgfonlayer{#1}\begingroup
    \aftergroup\endpgfonlayer
    \aftergroup\endgroup
  },
  /tikz/node on layer/.code={
    \gdef\node@@on@layer{%
      \setbox\tikz@tempbox=\hbox\bgroup\pgfonlayer{#1}\unhbox\tikz@tempbox\endpgfonlayer\egroup}
    \aftergroup\node@on@layer
  },
  /tikz/end node on layer/.code={
    \endpgfonlayer\endgroup\endgroup
  }
}

\def\node@on@layer{\aftergroup\node@@on@layer}

\makeatother
\begin{tikzpicture}[scale=2.2,tdplot_main_coords]
\tikzset{>=latex}
  % variables
  \def\rvec{.8}
  \def\thetavec{30}
  \def\phivec{60}

  \def\jointdist{0.4}
  % axes
  \coordinate (O) at (0,0,0);
  \draw[very thick,->] (0,0,0) -- (.5,0,0) node[distance from end=-0.3cm]{$\vecframe{x}{\earth{1}}{}$};
  \draw[very thick,->] (0,0,0) -- (0,.5,0) node[distance from end=-0.2cm]{$\vecframe{y}{\earth{1}}{}$};
  \draw[very thick,->] (0,0,0) -- (0,0,.5) node[distance from end=-0.2cm]{$\vecframe{z}{\earth{1}}{}\,\,\,$};

\pgfmathsetmacro{\cubex}{0.8}
\pgfmathsetmacro{\cubey}{0.2}
\pgfmathsetmacro{\cubez}{0.2}

\coordinate (OC) at (1,0.5,1);

% BOX 1
\tdplotsetrotatedcoordsorigin{(OC)}
\tdplotsetrotatedcoords{90}{-10}{0}
% \node[draw=none,shape=circle,fill, inner sep=1pt,tdplot_rotated_coords] at (OC){};
\draw[black,fill=green!40,opacity=1,tdplot_rotated_coords] ($(OC)+(\cubex/2,\cubey/2,\cubez/2)$) -- ++(-\cubex,0,0) -- ++(0,-\cubey,0) -- ++(\cubex,0,0) -- cycle;
\draw[black,fill=black!20,opacity=0,tdplot_rotated_coords] ($(OC)+(\cubex/2,\cubey/2,\cubez/2)$) -- ++(0,0,-\cubez) -- ++(0,-\cubey,0) -- ++(0,0,\cubez) -- cycle;
\draw[black,fill=black!10,opacity=0,tdplot_rotated_coords] ($(OC)+(\cubex/2,\cubey/2,\cubez/2)$) -- ++(-\cubex,0,0) -- ++(0,0,-\cubez) -- ++(\cubex,0,0) -- cycle;
\draw[black,fill=black!10,opacity=1,tdplot_rotated_coords] ($($(OC)+(\cubex/2,\cubey/2,\cubez/2)$)-(\cubex,\cubey,\cubez)$) -- ++(\cubex,0,0) -- ++(0,0,\cubez) -- ++(-\cubex,0,0) -- cycle;
\draw[black,fill=black!20,opacity=0,tdplot_rotated_coords] ($($(OC)+(\cubex/2,\cubey/2,\cubez/2)$)-(\cubex,\cubey,\cubez)$) -- ++(0,\cubey,0) -- ++(\cubex,0,0) -- ++(0,-\cubey,0) -- cycle;
\draw[black,fill=black!10,opacity=1,tdplot_rotated_coords] ($($(OC)+(\cubex/2,\cubey/2,\cubez/2)$)-(\cubex,\cubey,\cubez)$) -- ++(0,\cubey,0) -- ++(0,0,\cubez) -- ++(0,-\cubey,0) -- cycle;

% BOX 1 - BOX 2
\tdplotsetrotatedcoords{90}{-10}{0}
\tdplottransformrotmain{\cubex/2}{0}{0}
\coordinate (end_1) at ($(\tdplotresx, \tdplotresy, \tdplotresz)+(OC)$);
\tdplottransformrotmain{0.75*\cubex}{0}{0}
\coordinate (joint_1) at ($(\tdplotresx, \tdplotresy, \tdplotresz)+(OC)$);
\draw[line width=2pt,on layer=back] (end_1) -- (joint_1);
\node[circle,draw=black, line width=0.25mm, inner sep=3pt,fill=white,node on layer=front] at (joint_1){};

% disturbed box
\tdplotsetrotatedcoords{150}{-10}{0}
\tdplotsetrotatedcoordsorigin{(joint_1)}
\tdplottransformrotmain{-0.25*\cubex}{0}{0}
\coordinate (start_dist) at ($(\tdplotresx, \tdplotresy, \tdplotresz)+(joint_1)$);
\draw[] (start_dist) node[draw=none,shape=circle,fill, inner sep=1pt]{};
\tdplottransformrotmain{-0.75*\cubex}{0}{0}
\coordinate (OC_dist) at ($(\tdplotresx, \tdplotresy, \tdplotresz)+(joint_1)$);
\tdplotsetrotatedcoordsorigin{(OC_dist)}

\draw[orange,fill=black!20,opacity=1,tdplot_rotated_coords,on layer=front] ($(OC_dist)+(\cubex/2,\cubey/2,\cubez/2)$) -- ++(-\cubex,0,0) -- ++(0,-\cubey,0) -- ++(\cubex,0,0) -- cycle;
\draw[black,fill=black!0,opacity=0,tdplot_rotated_coords,dashed] ($(OC_dist)+(\cubex/2,\cubey/2,\cubez/2)$) -- ++(0,0,-\cubez) -- ++(0,-\cubey,0) -- ++(0,0,\cubez) -- cycle;
\draw[orange,fill=black!10,opacity=1,tdplot_rotated_coords] ($(OC_dist)+(\cubex/2,\cubey/2,\cubez/2)$) -- ++(-\cubex,0,0) -- ++(0,0,-\cubez) -- ++(\cubex,0,0) -- cycle;
\draw[black,fill=black!0,opacity=0,tdplot_rotated_coords,dashed] ($($(OC_dist)+(\cubex/2,\cubey/2,\cubez/2)$)-(\cubex,\cubey,\cubez)$) -- ++(\cubex,0,0) -- ++(0,0,\cubez) -- ++(-\cubex,0,0) -- cycle;
\draw[black,fill=black!0,opacity=0,tdplot_rotated_coords,dashed] ($($(OC_dist)+(\cubex/2,\cubey/2,\cubez/2)$)-(\cubex,\cubey,\cubez)$) -- ++(0,\cubey,0) -- ++(\cubex,0,0) -- ++(0,-\cubey,0) -- cycle;
\draw[orange,fill=black!10,opacity=1,tdplot_rotated_coords] ($($(OC_dist)+(\cubex/2,\cubey/2,\cubez/2)$)-(\cubex,\cubey,\cubez)$) -- ++(0,\cubey,0) -- ++(0,0,\cubez) -- ++(0,-\cubey,0) -- cycle;

\draw[line width=2pt,on layer=back,orange] (start_dist) -- (joint_1);

% delta stuff

% Line for Box 1
\tdplotsetrotatedcoords{90}{-10}{0}
\tdplotsetrotatedcoordsorigin{(OC)}
\tdplottransformrotmain{0.75*\cubex}{0}{0}
\coordinate (joint_height) at ($(0,0,\tdplotresz) + (0,0,1)$);

\coordinate (joint_proj) at ($(\tdplotresx,\tdplotresy,0)+(1,0.5,0)$);
% \draw[on layer=front] (joint_proj) node[draw=none,shape=circle,fill,color=red, inner sep=1pt]{};
\tdplottransformrotmain{-0.5*\cubex}{0}{0}
\coordinate (B1_front_proj) at ($(\tdplotresx,\tdplotresy,0)+(1,0.5,0)$);
\coordinate (B1_front) at ($(\tdplotresx,\tdplotresy,\tdplotresz)+(OC)$);
% \draw[on layer=front] (B1_front_proj) node[draw=none,shape=circle,fill,color=red, inner sep=1pt]{};
\draw[dotted, on layer=back] (joint_1) -- (joint_proj);
\draw[dotted, on layer=back] (joint_proj) -- (B1_front_proj);
\draw[dotted, on layer=back] (B1_front) -- (B1_front_proj);

\tdplotsetrotatedcoords{150}{-10}{0}
\tdplotsetrotatedcoordsorigin{(joint_1)}
\tdplottransformrotmain{-1.25*\cubex}{0}{0}
\coordinate (B2_end) at ($(\tdplotresx,\tdplotresy,\tdplotresz)+(joint_1)$);
\coordinate (B2_end_proj) at ($(\tdplotresx,\tdplotresy,0)+(joint_1)-(joint_height)$);
% \draw[on layer=front] (B2_end_proj) node[draw=none,shape=circle,fill,color=red, inner sep=1pt]{};

\draw[dotted, on layer=back] (B2_end) -- (B2_end_proj);
\draw[dotted, on layer=back] (joint_proj) -- (B2_end_proj);

% \tdplotsetrotatedcoordsorigin{(OC_dist)}
% \tdplottransformrotmain{0.75*\cubex}{0}{0}
% \draw[on layer=front] ($(\tdplotresx,\tdplotresy,0)+(OC_dist)-(0,0,1)$) node[draw=none,shape=circle,fill,color=red, inner sep=1pt]{};

\tdplotsetrotatedcoordsorigin{(joint_proj)}
\tdplotsetrotatedcoords{150}{0}{0}
\tdplotdrawarc[<-,tdplot_rotated_coords,orange,thick]{(0,0,0)}{1}{120}{180}{anchor=north}{$\delta$}

% coordinate system
\tdplotsetrotatedcoordsorigin{(OC)}
\tdplotsetrotatedcoords{90}{-10}{0}
\tdplottransformrotmain{0}{0}{\cubez/2}
\coordinate (cs_1) at ($(\tdplotresx, \tdplotresy, \tdplotresz)+(OC)$);
\tdplotsetrotatedcoordsorigin{(cs_1)}
\draw[thick,->,tdplot_rotated_coords] (cs_1) -- (-0.15,0,0) node [distance from end=-0.1cm]{};
\draw[thick,->,tdplot_rotated_coords] (cs_1) -- (0,-0.15,0) node[distance from end=-0.1cm]{};
\draw[thick,->,tdplot_rotated_coords] (cs_1) -- (0,0,0.15) node[distance from end=-0.1cm]{};
\node[above right = 0.1cm and -0.17cm] at (cs_1){\small{$\seg{2}$}};
\tdplotsetrotatedcoordsorigin{(OC)}
% ---------------------------------------------------------------------------------------------%
% BOX 2
\tdplotsetrotatedcoords{60}{10}{0} % ORIENTATION OF BOX 2
\tdplotsetrotatedcoordsorigin{(joint_1)}
\tdplottransformrotmain{0.25*\cubex}{0}{0}
\coordinate (start_2) at ($(\tdplotresx, \tdplotresy, \tdplotresz)+(joint_1)$);

\node[above left = -0.1cm and 0.1cm of joint_1] {$\joint{1}{2}$};

\draw[line width=2pt,on layer=middle] (joint_1) -- (start_2);
\tdplottransformrotmain{0.75*\cubex}{0}{0}
\coordinate (OC2) at ($(\tdplotresx, \tdplotresy, \tdplotresz)+(joint_1)$);
% \node[draw=none,shape=circle,fill, inner sep=1pt,color=red] at (start_2){};
% \node[draw=none,shape=circle,fill, inner sep=1pt,color=red] at (OC2){};

\tdplotsetrotatedcoordsorigin{(OC2)}

\draw[black,fill=red!40,opacity=1,tdplot_rotated_coords] ($(OC2)+(\cubex/2,\cubey/2,\cubez/2)$) -- ++(-\cubex,0,0) -- ++(0,-\cubey,0) -- ++(\cubex,0,0) -- cycle;
\draw[black,fill=black!20,opacity=0,tdplot_rotated_coords] ($(OC2)+(\cubex/2,\cubey/2,\cubez/2)$) -- ++(0,0,-\cubez) -- ++(0,-\cubey,0) -- ++(0,0,\cubez) -- cycle;
\draw[black,fill=black!20,opacity=0,tdplot_rotated_coords] ($(OC2)+(\cubex/2,\cubey/2,\cubez/2)$) -- ++(-\cubex,0,0) -- ++(0,0,-\cubez) -- ++(\cubex,0,0) -- cycle;
\draw[black,fill=black!10,opacity=1,tdplot_rotated_coords] ($($(OC2)+(\cubex/2,\cubey/2,\cubez/2)$)-(\cubex,\cubey,\cubez)$) -- ++(\cubex,0,0) -- ++(0,0,\cubez) -- ++(-\cubex,0,0) -- cycle;
\draw[black,fill=black!20,opacity=0,tdplot_rotated_coords] ($($(OC2)+(\cubex/2,\cubey/2,\cubez/2)$)-(\cubex,\cubey,\cubez)$) -- ++(0,\cubey,0) -- ++(\cubex,0,0) -- ++(0,-\cubey,0) -- cycle;
\draw[black,fill=black!10,opacity=1,tdplot_rotated_coords,on layer=back] ($($(OC2)+(\cubex/2,\cubey/2,\cubez/2)$)-(\cubex,\cubey,\cubez)$) -- ++(0,\cubey,0) -- ++(0,0,\cubez) -- ++(0,-\cubey,0) -- cycle;

% coordinate system
\tdplottransformrotmain{0}{0}{\cubez/2}
\coordinate (cs_2) at ($(\tdplotresx, \tdplotresy, \tdplotresz)+(OC2)$);
\tdplotsetrotatedcoordsorigin{(cs_2)}
\draw[thick,->,tdplot_rotated_coords,on layer=front] (cs_2) -- (-0.15,0,0) node [distance from end=-0.1cm]{};
\draw[thick,->,tdplot_rotated_coords,on layer=front] (cs_2) -- (0,-0.15,0) node[distance from end=-0.1cm]{};
\draw[thick,->,tdplot_rotated_coords,on layer=front] (cs_2) -- (0,0,0.15) node[distance from end=-0.1cm]{};
\node[above right=-00cm and 0.1cm] at (cs_2){\small{$\seg{1}$}};
\tdplotsetrotatedcoordsorigin{(OC2)}

% Orientations
\node[above right = 0.0cm and 1cm of OC2] {$\quatsegearth{1}{1}$};
\node[below left = 0.0cm and 0.5cm of OC] {$\quatsegearth{2}{1}$};
\node[below right = 0.0cm and 1.0cm of OC_dist,orange] {$\quatsegearth{2}{2}$};
\node[left = 0.0cm of O] {$\earth{1}$};

\node[above right = 0.1cm and 0.1cm of OC_dist] (A1) {};
\node[above right = .5cm of A1] (A2) {};
\draw[on layer=back] (A1) -- (A2);
\node[above right = -0.5cm of A2,align=left] {estimated\\orientation in $\earth{2}$};

\node[above left = 0.1cm and 0.1cm of OC] (B1) {};
\node[above left = 1cm of B1] (B2) {};
\draw[on layer=back] (B1) -- (B2);
\node[above left = 0cm and -0.5cm of B2,anchor=east,align=left] {true orientation in $\earth{1}$};

\end{tikzpicture} 

%% file: img/experiments/example_delta_3D_1.tex
\begin{tikzpicture}

\begin{axis}[%
width=0.951\figW,
height=\figH,
at={(0\figW,0\figH)},
scale only axis,
xmin=0,
xmax=6.28318530717959,
xtick={0,0.785398163397448,1.5707963267949,2.35619449019234,3.14159265358979,3.92699081698724,4.71238898038469,5.49778714378214,6.28318530717959},
xticklabels={{$0$},{$\frac{\pi}{4}$},{$\frac{\pi}{2}$},{$\frac{3\pi}{4}$},{$\pi$},{$\frac{5\pi}{4}$},{$\frac{3\pi}{2}$},{$\frac{7\pi}{4}$},{$2\pi$}},
xlabel style={font=\color{white!15!black}},
ymin=5,
ymax=45,
y label style={at={(axis description cs:0.4,.5)},anchor=south},
ylabel={$c(w,\heest)$},
axis background/.style={fill=white},
xmajorgrids,
ymajorgrids,
grid style={dotted}
]
\addplot [color=black, line width=1.0pt, forget plot]
  table[row sep=crcr]{%
0	43.8222222222222\\
2.04203522483336	41.2222222222222\\
2.07694180987325	39.1777777777778\\
2.09439510239319	36.1555555555556\\
2.11184839491314	32.1333333333333\\
2.12930168743308	27.1111111111111\\
2.14675497995302	25.0888888888889\\
2.16420827247297	17.0666666666667\\
2.18166156499291	11.0444444444444\\
2.19911485751285	7.02222222222223\\
2.2165681500328	6\\
2.23402144255274	16.0222222222222\\
2.25147473507268	17.0444444444444\\
2.26892802759263	20.0666666666667\\
2.28638132011258	25.0888888888889\\
2.30383461263251	26.1111111111111\\
2.3387411976724	26.1555555555556\\
2.39110107523223	29.2222222222222\\
2.42600766027212	29.2666666666667\\
2.44346095279206	34.2888888888889\\
2.460914245312	38.3111111111111\\
2.49582083035189	38.3555555555556\\
2.53072741539178	40.4\\
2.56563400043166	40.4444444444444\\
2.58308729295161	41.4666666666667\\
5.35816080362259	45\\
6.26573201465964	43.8444444444444\\
};
\end{axis}
\end{tikzpicture}% 

%% file: img/experiments/example_delta_3D_3.tex
\begin{tikzpicture}

\begin{axis}[%
width=0.951\figW,
height=\figH,
at={(0\figW,0\figH)},
scale only axis,
xmin=0,
xmax=6.28318530717959,
xtick={0,0.785398163397448,1.5707963267949,2.35619449019234,3.14159265358979,3.92699081698724,4.71238898038469,5.49778714378214,6.28318530717959},
xticklabels={{$0$},{$\frac{\pi}{4}$},{$\frac{\pi}{2}$},{$\frac{3\pi}{4}$},{$\pi$},{$\frac{5\pi}{4}$},{$\frac{3\pi}{2}$},{$\frac{7\pi}{4}$},{$2\pi$}},
xlabel style={font=\color{white!15!black}},
ymin=0,
ymax=45,
ytick={0,5,10,15,20,25,30,35,40,45},
yticklabels={\empty},
axis background/.style={fill=white},
xmajorgrids,
ymajorgrids,
grid style={dotted}
]
\addplot [color=black, line width=1.0pt, forget plot]
  table[row sep=crcr]{%
0	44.3333333333333\\
1.8151424220741	42.0222222222222\\
1.83259571459405	41\\
2.02458193231342	40.7555555555556\\
2.07694180987325	37.6888888888889\\
2.25147473507268	37.4666666666667\\
2.28638132011258	33.4222222222222\\
2.30383461263251	32.4\\
2.3387411976724	32.3555555555556\\
2.37364778271229	30.3111111111111\\
2.39110107523223	30.2888888888889\\
2.40855436775217	29.2666666666667\\
2.42600766027212	29.2444444444444\\
2.44346095279206	28.2222222222222\\
2.460914245312	24.2\\
2.47836753783195	23.1777777777778\\
2.49582083035189	19.1555555555556\\
2.51327412287183	17.1333333333333\\
2.53072741539178	14.1111111111111\\
2.54818070791172	14.0888888888889\\
2.56563400043166	13.0666666666667\\
2.58308729295161	11.0444444444444\\
2.60054058547155	7.02222222222223\\
2.61799387799149	2\\
2.63544717051144	1.02222222222222\\
2.65290046303138	2.04444444444444\\
2.68780704807127	2.08888888888889\\
2.70526034059121	3.11111111111111\\
2.72271363311116	5.13333333333333\\
2.74016692563109	5.15555555555556\\
2.75762021815104	6.17777777777778\\
2.79252680319092	10.2222222222222\\
2.80998009571087	11.2444444444444\\
2.82743338823082	13.2666666666667\\
2.84488668075075	13.2888888888889\\
2.87979326579065	15.3333333333333\\
2.89724655831058	17.3555555555556\\
2.91469985083053	18.3777777777778\\
2.94960643587041	24.4222222222222\\
2.96705972839036	31.4444444444444\\
2.98451302091031	33.4666666666667\\
3.00196631343024	33.4888888888889\\
3.01941960595019	34.5111111111111\\
3.10668606854991	34.6222222222222\\
3.12413936106985	35.6444444444444\\
3.1415926535898	35.6666666666667\\
3.15904594610974	38.6888888888889\\
3.17649923862968	39.7111111111111\\
3.24631240870945	39.8\\
3.28121899374934	41.8444444444444\\
5.75958653158128	45\\
6.26573201465964	44.3555555555556\\
};
\end{axis}
\end{tikzpicture}% 

%% file: img/experiments/example_delta_3D_2.tex
\begin{tikzpicture}

\begin{axis}[%
width=0.951\figW,
height=\figH,
at={(0\figW,0\figH)},
scale only axis,
xmin=0,
xmax=6.28318530717959,
xtick={0,0.785398163397448,1.5707963267949,2.35619449019234,3.14159265358979,3.92699081698724,4.71238898038469,5.49778714378214,6.28318530717959},
xticklabels={{$0$},{$\frac{\pi}{4}$},{$\frac{\pi}{2}$},{$\frac{3\pi}{4}$},{$\pi$},{$\frac{5\pi}{4}$},{$\frac{3\pi}{2}$},{$\frac{7\pi}{4}$},{$2\pi$}},
xlabel style={font=\color{white!15!black}},
ymin=0,
ymax=14,
ytick={0,2,4,6,8,10,12,14},
yticklabels={\empty},
axis background/.style={fill=white},
xmajorgrids,
ymajorgrids,
grid style={dotted}
]
\addplot [color=black, line width=1.0pt, forget plot]
  table[row sep=crcr]{%
0	12.4222222222222\\
1.51843644923507	10.4888888888889\\
1.53588974175501	9.46666666666667\\
1.79768912955416	9.13333333333333\\
1.8151424220741	4.11111111111111\\
1.86750229963393	4.04444444444444\\
1.90240888467382	2\\
2.09439510239319	2.24444444444445\\
2.11184839491314	1.26666666666667\\
2.40855436775217	1.64444444444444\\
2.42600766027212	2.66666666666667\\
2.51327412287183	2.77777777777778\\
2.53072741539178	3.8\\
2.65290046303138	3.95555555555556\\
2.67035375555133	4.97777777777778\\
2.68780704807127	9\\
2.72271363311115	9.04444444444444\\
2.7401669256311	10.0666666666667\\
2.75762021815104	10.0888888888889\\
2.77507351067098	11.1111111111111\\
5.04400153826361	14\\
6.26573201465964	12.4444444444444\\
};
\end{axis}
\end{tikzpicture}% 

%% file: img/experiments/MechanicalModel_3D_angle.pdf_tex
%% Accompanies image file 'MechanicalModel_3D_angle.pdf' (pdf, eps, ps)
\begingroup%
  \makeatletter%
  \providecommand\color[2][]{%
    \errmessage{(Inkscape) Color is used for the text in Inkscape, but the package 'color.sty' is not loaded}%
    \renewcommand\color[2][]{}%
  }%
  \providecommand\transparent[1]{%
    \errmessage{(Inkscape) Transparency is used (non-zero) for the text in Inkscape, but the package 'transparent.sty' is not loaded}%
    \renewcommand\transparent[1]{}%
  }%
  \providecommand\rotatebox[2]{#2}%
  \newcommand*\fsize{\dimexpr\f@size pt\relax}%
  \newcommand*\lineheight[1]{\fontsize{\fsize}{#1\fsize}\selectfont}%
  \ifx\svgwidth\undefined%
    \setlength{\unitlength}{1512.70733643bp}%
    \ifx\svgscale\undefined%
      \relax%
    \else%
      \setlength{\unitlength}{\unitlength * \real{\svgscale}}%
    \fi%
  \else%
    \setlength{\unitlength}{\svgwidth}%
  \fi%
  \global\let\svgwidth\undefined%
  \global\let\svgscale\undefined%
  \makeatother%
  \begin{picture}(1,0.2765173)%
    \lineheight{1}%
    \setlength\tabcolsep{0pt}%
    \put(0,0){\includegraphics[width=\unitlength,page=1]{MechanicalModel_3D_angle.pdf}}%
  \end{picture}%
\endgroup%

%% file: img/experiments/log_plot_1.tex
\definecolor{mycolor1}{rgb}{0.00000,0.44700,0.74100}%
\definecolor{mycolor2}{rgb}{0.85000,0.32500,0.09800}%
\definecolor{mycolor3}{rgb}{0.92900,0.69400,0.12500}%
\definecolor{mycolor4}{rgb}{0.49400,0.18400,0.55600}%
\definecolor{mycolor5}{rgb}{0.46600,0.67400,0.18800}%
\definecolor{mycolor6}{rgb}{0.30100,0.74500,0.93300}%
\definecolor{mycolor7}{rgb}{0.63500,0.07800,0.18400}%
\begin{tikzpicture}

\begin{axis}[%
width=0.951\figW,
height=\figH,
at={(0\figW,0\figH)},
scale only axis,
xmin=0,
xmax=10,
xlabel style={font=\color{white!15!black}},
xlabel={Time $[\s]$},
ymode=log,
ymin=0.5,
ymax=100,
yminorticks=true,
ylabel style={font=\color{white!15!black}},
ylabel={Angle Error $\epsilon\,[\g]$},
y label style={at={(axis description cs:0.22,.5)},anchor=south},
axis background/.style={fill=white},
axis x line*=bottom,
axis y line*=left,
xmajorgrids,
ymajorgrids,
yminorgrids,
grid style={dotted}
]
\addplot [color=mycolor1, forget plot, line width = 0.6pt]
  table[row sep=crcr]{%
0	34.8905457679974\\
0.24	34.8152915637126\\
0.32	33.0722375050204\\
0.380000000000001	30.5844783829797\\
0.44	28.1058160215885\\
0.48	26.4564067008774\\
0.5	25.6396379073771\\
0.52	23.4321645989772\\
0.540000000000001	21.2491426655741\\
0.56	20.2178570875355\\
0.58	19.9362006820497\\
0.6	19.0542724178555\\
0.619999999999999	17.238201367815\\
0.640000000000001	15.9403337546391\\
0.68	14.6727249382661\\
0.699999999999999	14.1479221769496\\
0.720000000000001	13.9349044195317\\
0.74	13.9349044195317\\
0.76	13.7856553992816\\
0.800000000000001	13.1200213691873\\
0.82	12.8937286793616\\
0.84	12.7328404847742\\
0.859999999999999	12.5378272309973\\
0.880000000000001	12.4079197454318\\
0.9	12.3233731901079\\
0.92	12.101314247458\\
0.94	11.7939373169854\\
0.960000000000001	11.6497306000745\\
1.16	11.6497306000745\\
1.18	11.6233522828857\\
1.22	11.3365648640094\\
1.24	11.2803345406928\\
1.26	11.1178649990344\\
1.28	10.9015021329325\\
1.32	10.7408264614955\\
1.34	10.6049171180232\\
1.36	10.5319130580398\\
1.42	10.521681893373\\
1.64	10.521681893373\\
1.66	10.4395779884953\\
1.68	10.1751743601048\\
1.7	9.68814798683044\\
1.72	9.19625910156962\\
1.76	8.63171947026138\\
1.8	7.95463837710974\\
1.82	7.8120197518717\\
1.96	7.8120197518717\\
1.98	7.77772506234355\\
2	7.52979293325846\\
2.02	7.11992538687323\\
2.04	6.87617243081322\\
2.06	6.79263312113757\\
2.08	6.61403081996173\\
2.1	6.25752915642813\\
2.12	5.89834998641641\\
2.14	5.75308663920635\\
2.18	5.75308663920635\\
2.2	5.68452263933697\\
2.22	5.58044081984725\\
2.68	5.58044081984725\\
2.7	5.54346434772088\\
2.76	5.54346434772088\\
2.78	5.55042592142965\\
2.82	5.75261046265317\\
2.84	5.77449234174946\\
2.86	5.849306300281\\
2.88	5.91257519097842\\
3.24	5.91257519097842\\
3.28	5.80118142814661\\
3.32	5.75065516225957\\
3.48	5.750158514372\\
3.54	5.69629808921641\\
3.56	5.61647867585082\\
3.58	5.55592884263234\\
3.64	5.54200569521485\\
3.66	5.50529279824868\\
3.68	5.43629996091297\\
3.72	5.25997841369343\\
3.74	5.19963934564848\\
3.78	5.11920483899204\\
3.8	5.04215661009253\\
3.82	4.98198047931823\\
4.2	4.98198047931823\\
4.24	4.8732724199067\\
4.26	4.81373678371014\\
4.32	4.44590502999185\\
4.34	4.22874110920222\\
4.36	3.9279895664128\\
4.38	3.62055964530982\\
4.4	3.40642616711413\\
4.42	3.31278788887754\\
4.44	3.26434451643073\\
4.46	3.15978981112263\\
4.48	3.02574055302651\\
4.5	2.9495297459248\\
4.54	2.84910137557211\\
4.56	2.82971737437805\\
4.58	2.82009470233792\\
4.6	2.80190459199707\\
4.62	2.77802333640587\\
4.64	2.743296851213\\
4.68	2.65915579049024\\
4.7	2.63298857973501\\
4.72	2.61720852081795\\
4.74	2.59050156775542\\
4.76	2.55673859725508\\
4.8	2.47501130496547\\
4.86	2.36392441694879\\
4.88	2.34676381858276\\
4.9	2.31460980363041\\
4.92	2.26202180460031\\
4.94	2.22277630760654\\
4.98	2.1887599278914\\
5.02	2.16368108079915\\
5.04	2.14643628410887\\
5.06	2.13764490021285\\
5.08	2.12086018169284\\
5.12	2.04078154496449\\
5.14	2.00771895171382\\
5.16	1.99049072151475\\
5.64	1.99049072151475\\
5.66	2.03879396086344\\
5.68	2.18399437574601\\
5.74	2.83030480754608\\
5.76	3.0855349297514\\
5.78	3.2170521002548\\
5.8	3.22509681127266\\
6.42	3.22754808115381\\
6.44	3.21394153197117\\
6.48	3.13456187668652\\
6.52	3.10023886192276\\
6.54	3.06502573435893\\
6.6	2.89789412753059\\
6.64	2.82613157292446\\
6.66	2.76935262918021\\
6.68	2.7296803938664\\
6.7	2.71780071233294\\
6.74	2.70557870379456\\
6.76	2.67860991157921\\
6.78	2.64436471899635\\
6.8	2.62571838473415\\
6.82	2.5998346866261\\
6.84	2.5565026717079\\
6.86	2.53002343125751\\
6.88	2.51544922960301\\
6.9	2.49025536081748\\
6.92	2.45933447774155\\
6.94	2.38870205743673\\
6.96	2.20180452858157\\
6.98	2.06200030990241\\
7	2.05395105036137\\
7.06	2.05395105036137\\
7.1	2.03939433141812\\
7.12	2.03939433141812\\
7.14	2.03459116693104\\
7.64	2.03637705231486\\
7.68	2.05743156871446\\
7.7	2.08052047645586\\
7.72	2.09855854168723\\
7.82	2.16062230317386\\
7.84	2.17561719277328\\
7.9	2.20345704722253\\
7.92	2.21044573509986\\
7.94	2.22545821008836\\
7.96	2.27586536729965\\
7.98	2.34277842137916\\
8	2.37208067510369\\
8.06	2.37616755306794\\
8.08	2.40075761977443\\
8.1	2.43638895016355\\
8.14	2.47640448139286\\
8.16	2.52138829826972\\
8.18	2.54907626929121\\
8.66	2.54907626929121\\
8.68	2.52138829826972\\
8.7	2.45337085274794\\
8.72	2.18969594413256\\
8.74	1.72171047379021\\
8.76	1.41170524505872\\
8.78	1.32541171718627\\
8.8	1.29559811596014\\
8.82	1.27890267599379\\
8.86	1.25723447043897\\
9.36	1.25933116429201\\
9.38	1.27890267599379\\
9.4	1.29559811596014\\
9.44	1.31858852876644\\
9.46	1.32668700007077\\
9.78	1.32668700007077\\
9.86	1.63637678885129\\
9.9	1.74513903050247\\
9.94	1.81237799903262\\
9.98	1.81381020691944\\
10	1.81958892271551\\
};
\addplot [color=mycolor2, forget plot, line width = 0.6pt]
  table[row sep=crcr]{%
0	19.6149367506836\\
0.32	20.5919020987223\\
0.44	20.8857350633443\\
0.56	21.0782405193538\\
0.68	21.0782405193538\\
0.699999999999999	21.1708291852939\\
0.76	21.6750375878053\\
0.84	22.489003801821\\
0.880000000000001	22.6718425922375\\
0.92	22.8000607006251\\
0.960000000000001	22.8586199245851\\
1.14	22.8910314368274\\
1.66	22.8613755951246\\
1.68	22.8078149078563\\
1.7	22.6595636461353\\
1.72	22.366631497536\\
1.74	21.701943825801\\
1.78	19.7365205318992\\
1.8	19.2515834523302\\
2.08	19.2515834523302\\
2.1	18.9275843257223\\
2.12	18.3587995676728\\
2.14	17.7527741997931\\
2.16	17.3856772797029\\
2.28	17.3798200960506\\
2.42	17.3798200960506\\
2.44	16.7259454926575\\
2.46	15.2533016056922\\
2.48	14.0908641369683\\
2.5	13.4861200766329\\
2.52	12.9633106955222\\
2.54	12.3778516258584\\
2.56	11.4950603355447\\
2.58	10.5901608749727\\
2.6	9.96123854860132\\
2.62	9.1787575177113\\
2.64	8.3679848279276\\
2.66	8.0566045766573\\
2.72	8.0566045766573\\
2.74	7.73193899337737\\
2.76	6.85939681067145\\
2.78	5.81730535602859\\
2.8	4.84093635864742\\
2.82	4.06835764762464\\
2.84	3.70463702992559\\
2.86	3.57012278621451\\
2.88	3.46174409600329\\
2.9	3.39398821128965\\
2.92	3.3733928219167\\
2.94	3.3733928219167\\
2.96	3.34855498669403\\
3	3.2183547067113\\
3.02	3.16258285726723\\
3.04	3.08052032818683\\
3.06	3.00796152452669\\
3.08	2.9656146068189\\
3.12	2.86855514505356\\
3.14	2.84021755891999\\
3.16	2.80235822546738\\
3.18	2.75533768685669\\
3.2	2.7191649756067\\
3.22	2.67453829136722\\
3.24	2.60328326516286\\
3.26	2.50770611944574\\
3.28	2.42221500034741\\
3.3	2.37897754187247\\
3.36	2.32793467270374\\
3.38	2.29132616780162\\
3.4	2.24935781459037\\
3.42	2.22622420290457\\
3.44	2.21276775213173\\
3.46	2.19056439693964\\
3.48	2.17780171252479\\
3.58	2.17646927791536\\
3.64	2.16558295717429\\
3.68	2.16558295717429\\
3.7	2.16002577209991\\
3.72	2.11874541476615\\
3.74	2.04497911201579\\
3.76	1.99347064499297\\
3.78	1.98000530846115\\
3.8	1.98000530846115\\
3.82	1.969016031895\\
3.84	1.94862587156976\\
3.86	1.92084787008893\\
3.88	1.86098115882352\\
3.9	1.77600940773128\\
3.92	1.71585395450763\\
3.94	1.69233673193928\\
3.96	1.6650502041457\\
4	1.62493725544923\\
4.02	1.58836823743942\\
4.04	1.57142595674383\\
4.72	1.57138234626188\\
4.78	1.59551876007387\\
4.8	1.60480942808539\\
4.88	1.65618837429521\\
4.9	1.67570349741805\\
4.92	1.72671587124041\\
4.94	1.83709907567199\\
4.96	1.94407326947683\\
4.98	2.02673121113109\\
5	2.12261440071261\\
5.02	2.20089108769772\\
5.04	2.24860172635183\\
5.06	2.28509302349725\\
5.08	2.30109904926601\\
5.18	2.30109904926601\\
5.22	2.3332913322876\\
5.78	2.33662980078694\\
5.88	2.35048918572827\\
6.06	2.34897991226124\\
6.1	2.33996826928627\\
6.16	2.33996826928627\\
6.2	2.33002198881912\\
6.24	2.32807709893492\\
6.26	2.32083225126013\\
6.28	2.30495601760125\\
6.3	2.29437974173294\\
6.34	2.26129326498499\\
6.4	2.25972970361168\\
6.44	2.24698890214105\\
6.5	2.21814973626608\\
6.52	2.20758408761705\\
6.54	2.18799673663779\\
6.58	2.11099786458832\\
6.6	2.09215223889683\\
6.64	2.06274068897939\\
6.66	2.03299214639432\\
6.68	1.99449785371749\\
6.7	1.96683460847128\\
6.72	1.95004108526265\\
6.74	1.94348944574598\\
6.78	1.94348944574598\\
6.82	1.92339234748633\\
6.92	1.92339234748633\\
6.96	1.90429886746403\\
7	1.90429886746403\\
7.08	1.8899751358196\\
7.18	1.88034395962892\\
7.22	1.86476864380863\\
7.32	1.79504473713808\\
7.34	1.77652432532273\\
7.36	1.7632240571199\\
7.4	1.74955105366226\\
7.48	1.73425189309459\\
7.5	1.71648626905604\\
7.54	1.65146549622899\\
7.56	1.6377389967314\\
7.58	1.6313673391111\\
7.64	1.62403853138925\\
7.72	1.62403853138925\\
7.74	1.62017185158478\\
7.8	1.58498295152202\\
7.82	1.5673393057166\\
7.84	1.5433155983008\\
7.86	1.52664758243827\\
7.88	1.52079273327324\\
7.9	1.49653680065403\\
7.92	1.45094850106937\\
7.94	1.41537318030216\\
7.96	1.38796413445006\\
8.46	1.38796413445006\\
8.48	1.41537318030216\\
8.5	1.43727152924318\\
8.54	1.46961136043739\\
8.56	1.4965320175836\\
8.58	1.51344804636017\\
8.64	1.52725477313553\\
8.72	1.52725477313553\\
8.78	1.5702661193433\\
8.8	1.60266561107296\\
8.82	1.63169846241322\\
8.86	1.67260741500924\\
8.88	1.69848429158846\\
8.9	1.71665249572935\\
8.94	1.73056605289311\\
8.96	1.75585023395031\\
8.98	1.80648432367119\\
9	1.8464101050466\\
9.02	1.86417376259862\\
9.06	1.88643079245222\\
9.14	1.95402008391595\\
9.16	1.95862346694183\\
9.2	1.95862346694183\\
9.22	1.97552011268939\\
9.26	2.05317272724615\\
9.28	2.08200282729865\\
9.3	2.10531244845418\\
9.32	2.1164608746302\\
9.34	2.12317927382002\\
9.38	2.14791084688222\\
9.46	2.17388575971305\\
9.56	2.1766459838821\\
9.6	2.17940620805115\\
9.66	2.17940620805115\\
9.74	2.1984730648811\\
9.84	2.21719273817404\\
9.92	2.21975859561173\\
9.96	2.2283601005508\\
10	2.2425137191088\\
};
\addplot [color=mycolor3, forget plot, line width = 0.6pt]
  table[row sep=crcr]{%
0	19.176100296264\\
0.34	18.7667964400143\\
0.74	18.7177128036957\\
0.960000000000001	18.4547178818507\\
1	18.4060846905142\\
1.02	17.7721217879699\\
1.04	15.9115214544953\\
1.06	13.5077501650621\\
1.08	11.812190365932\\
1.1	11.4398352534508\\
1.58	11.4398352534508\\
1.6	11.7424944197012\\
1.62	11.9263651291995\\
1.64	11.9604847026522\\
2.12	11.9604847026522\\
2.14	11.9263651291995\\
2.2	11.6113073578521\\
2.22	11.5775802850834\\
2.46	11.5775802850834\\
2.5	11.4675319841758\\
2.56	11.3453084951848\\
2.58	11.3057571532005\\
2.6	11.2374876587505\\
2.66	11.1148867121756\\
2.7	10.9772712872804\\
2.72	10.7428071964095\\
2.74	10.1670842246394\\
2.76	9.51037786301831\\
2.8	8.63758574299618\\
2.82	8.18944468998963\\
2.84	7.73473794808327\\
2.86	7.36924124218161\\
2.88	7.09113417396536\\
2.9	6.7827247934935\\
2.94	6.30168254215531\\
2.96	6.08245738630595\\
2.98	5.92533547013072\\
3.02	5.6553694332021\\
3.04	5.50716830316459\\
3.06	5.33883213565325\\
3.08	5.10709034905223\\
3.1	4.82589659355396\\
3.12	4.49419823065763\\
3.14	4.30821843070954\\
3.16	4.30821843070954\\
3.18	4.21240182246988\\
3.2	4.05962895012656\\
3.26	3.72461816903223\\
3.28	3.5745068723046\\
3.3	3.39842500248029\\
3.32	3.10686041924237\\
3.34	2.74929137471387\\
3.36	2.50346880626162\\
3.38	2.29031510087321\\
3.4	2.04939831923751\\
3.42	1.8905913576722\\
3.44	1.82889475816856\\
3.46	1.75384014172716\\
3.48	1.72311075498092\\
3.96	1.72311075498092\\
3.98	1.77040478920206\\
4	1.8011341759483\\
4.02	1.82719169435652\\
4.06	1.8635745216162\\
4.08	1.90568969110407\\
4.1	1.98185754196176\\
4.12	2.04775542913394\\
4.14	2.0773968991448\\
4.2	2.10351865772792\\
4.24	2.11453110817652\\
4.28	2.13817558822344\\
4.3	2.14736782912467\\
4.32	2.16330236033436\\
4.34	2.18393619521765\\
4.36	2.23631552817703\\
4.38	2.31630524586155\\
4.4	2.3640653467557\\
4.46	2.41643449615552\\
4.52	2.44641432148664\\
4.54	2.46253395015251\\
4.58	2.5064061607956\\
4.64	2.53836836128357\\
4.7	2.57270867700245\\
4.72	2.59483076297094\\
4.74	2.6529677697431\\
4.76	2.74303697395321\\
4.78	2.88908798767861\\
4.8	3.08636014949724\\
4.82	3.18421830298331\\
4.84	3.21464212617435\\
4.86	3.32234597175646\\
4.88	3.45290463522998\\
4.9	3.50618327631244\\
4.94	3.51206175321811\\
4.96	3.56366923242185\\
4.98	3.66990853480042\\
5	3.75637296417892\\
5.5	3.75637296417892\\
5.52	3.66789899408347\\
5.54	3.55578121479923\\
5.56	3.48928865156438\\
5.58	3.43857492259661\\
5.62	3.29340495334436\\
5.64	3.26255999335804\\
6.16	3.26255999335804\\
6.18	3.43738093027234\\
6.2	3.76362960294871\\
6.22	3.91505733871078\\
6.74	3.91505733871078\\
6.76	3.90598305343109\\
6.78	3.86834262191899\\
6.8	3.7948802945737\\
6.82	3.70265734693804\\
6.84	3.6553305804153\\
6.94	3.6553305804153\\
6.96	3.63685008106835\\
7.02	3.53040929355362\\
7.04	3.46744955453372\\
7.06	3.41585241043505\\
7.08	3.37933794566446\\
7.1	3.35722063969494\\
7.12	3.2812897315738\\
7.14	3.12671329064183\\
7.16	3.01295399504637\\
7.32	3.01243282904127\\
7.34	2.98730323304248\\
7.94	2.99231216234324\\
7.96	2.99744380578508\\
8	3.02680144403413\\
8.16	3.02680144403413\\
8.2	2.99744380578508\\
8.28	2.99034047795933\\
8.38	2.99428384672715\\
8.4	2.99744380578508\\
8.44	3.02680144403413\\
8.52	3.02877312841805\\
8.78	3.02877312841805\\
8.8	3.04784425119592\\
8.82	3.10804933934561\\
8.84	3.1780381505486\\
8.86	3.21289875700071\\
8.9	3.21890451762168\\
8.94	3.21890451762168\\
8.96	3.23173638846553\\
8.98	3.28892052580299\\
9	3.37184968626452\\
9.02	3.42072797644722\\
9.12	3.52584860631427\\
9.14	3.58917092371952\\
9.16	3.67098120031786\\
9.18	3.71383839767949\\
9.22	3.74898952117987\\
9.24	3.77134754990783\\
9.26	3.80608602671165\\
9.28	3.88265228664837\\
9.3	3.98892562739057\\
9.32	4.06643096469334\\
9.34	4.09137141996898\\
9.36	4.09137141996898\\
9.38	4.1120067403823\\
9.4	4.1468045380912\\
9.44	4.19345842926111\\
9.46	4.23650767092092\\
9.52	4.31945407314635\\
9.54	4.33541064695832\\
9.58	4.44024115502446\\
9.62	4.59186743515186\\
9.66	4.59466181139717\\
9.7	4.65225854863388\\
9.74	4.65225854863388\\
9.78	4.68628088488654\\
10	4.68628088488654\\
};
\addplot [color=mycolor4, forget plot, line width = 0.6pt]
  table[row sep=crcr]{%
0	19.1358141040651\\
0.140000000000001	19.2014493572886\\
0.42	19.3545152900655\\
0.540000000000001	19.4611576736117\\
0.56	19.4611576736117\\
0.6	19.6501989121185\\
0.66	19.6501989121185\\
0.68	19.8741918009834\\
0.699999999999999	20.4083537342788\\
0.720000000000001	20.7185227787092\\
0.74	20.8933408189384\\
1.22	20.8933408189384\\
1.24	20.3105304753122\\
1.26	18.5310538794001\\
1.28	17.330236115812\\
1.32	16.7365880353098\\
1.38	16.0971681873565\\
1.4	15.9658037940466\\
1.46	15.9471097494546\\
1.58	15.9219093155598\\
1.6	15.7981778657746\\
1.62	15.5613639551461\\
1.64	15.3646078390056\\
1.7	14.9576059695087\\
1.72	14.7336134748327\\
1.74	14.4575314371648\\
1.76	14.2279541409995\\
1.78	14.1483391981151\\
1.96	14.1295290488548\\
2	14.1107188995945\\
2.08	14.0967025093519\\
2.14	14.0727705016476\\
2.32	13.9748713971073\\
2.52	13.9703347436947\\
2.56	13.8114742962892\\
2.94	13.7953835159116\\
3.02	13.7390684997276\\
3.1	13.7017024778361\\
3.26	13.6230935527809\\
3.32	13.5496052007848\\
3.46	13.4426895441679\\
3.48	13.3982005148431\\
3.5	13.2868782986849\\
3.52	13.042660144993\\
3.54	12.7063502994484\\
3.58	11.9281791119124\\
3.6	11.6918138382206\\
3.62	11.3897987368298\\
3.64	10.9975931326927\\
3.66	10.4964220038955\\
3.68	9.64189177562193\\
3.7	8.81623369429257\\
3.72	8.31918737580726\\
3.74	7.76821088710618\\
3.78	6.45132579604559\\
3.8	5.89470695774447\\
3.82	5.27522694218646\\
3.84	4.83442521781337\\
3.86	4.58555257090028\\
3.9	3.99544263079475\\
3.92	3.86869911363458\\
3.94	3.83880336163399\\
3.98	3.71938515814967\\
4	3.69954850340822\\
4.02	3.69954850340822\\
4.04	3.6694342722923\\
4.06	3.60920885075854\\
4.08	3.56726074734527\\
4.1	3.54311910701694\\
4.12	3.49190076790833\\
4.14	3.34025077148996\\
4.16	3.10955963323124\\
4.18	2.95041074020891\\
4.2	2.89157680972845\\
4.22	2.82278123686253\\
4.24	2.72456359402166\\
4.28	2.58404016134811\\
4.3	2.43002787554382\\
4.32	2.22179104049927\\
4.34	2.08305240811653\\
4.36	2.04092880654017\\
4.6	2.04038131096617\\
4.64	2.0214070645632\\
4.74	2.0214070645632\\
4.86	1.96658119705417\\
5.34	1.96658119705417\\
5.36	1.95318534770306\\
5.38	1.9149763130163\\
5.42	1.81831797586044\\
5.46	1.7452592146351\\
5.48	1.69725353120024\\
5.5	1.66199774675388\\
5.56	1.59924229380848\\
5.58	1.57336475801513\\
5.6	1.54285493482215\\
5.62	1.5094597800186\\
5.64	1.48626501033289\\
5.68	1.47492251447097\\
5.7	1.47056916487435\\
6.18	1.47056916487435\\
6.24	1.48348720169883\\
6.26	1.49221638241669\\
6.28	1.50772380137312\\
6.3	1.54508483059152\\
6.34	1.66166477254651\\
6.36	1.69123385199844\\
6.38	1.69123385199844\\
6.42	1.75579160472629\\
6.5	1.75848847003304\\
6.52	1.75848847003304\\
6.54	1.76745192072548\\
6.56	1.80498036427723\\
6.58	1.8561081722151\\
6.6	1.88283206771339\\
6.78	1.9658129495961\\
6.8	1.98022318804269\\
6.88	2.06957894017966\\
6.9	2.0845764095862\\
6.94	2.1026352454359\\
6.96	2.1026352454359\\
6.98	2.11347381904148\\
7	2.13325402988418\\
7.02	2.16985124343961\\
7.04	2.22584602485784\\
7.06	2.26729739086946\\
7.08	2.29616334000337\\
7.1	2.35301154570421\\
7.12	2.39410596318283\\
7.14	2.39410596318283\\
7.18	2.46850825213985\\
7.2	2.5546468486172\\
7.22	2.77252783849045\\
7.24	2.97248412526032\\
7.26	3.06189019583784\\
7.3	3.22853453415669\\
7.32	3.25895108026947\\
7.92	3.25895108026947\\
7.96	3.22749565650265\\
8	3.21048218917406\\
8.14	3.21048218917406\\
8.16	3.20206858101369\\
8.18	3.16265936548813\\
8.2	3.13166375812293\\
8.22	3.11474341629826\\
8.24	3.07538922781588\\
8.28	3.03431312649434\\
8.3	2.99050670767252\\
8.32	2.9653425435145\\
8.34	2.92755658903179\\
8.36	2.86954091397174\\
8.4	2.81951468548947\\
8.42	2.80037098567057\\
8.5	2.79488222269167\\
8.52	2.759056185845\\
8.54	2.66368494019615\\
8.56	2.55179371421346\\
8.58	2.50312099086234\\
8.64	2.50130552171285\\
8.74	2.50130552171285\\
8.78	2.46780598167878\\
8.82	2.4659905125293\\
8.84	2.44041239821435\\
8.86	2.36297283581393\\
8.88	2.31111138772845\\
8.9	2.28441354403833\\
8.92	2.2154077017003\\
8.94	2.1730997030524\\
9.02	2.17197997367721\\
9.04	2.16443443892136\\
9.06	2.11635988666634\\
9.08	2.03456973130961\\
9.1	1.99442832282722\\
9.28	1.99442832282722\\
9.3	1.97434393040872\\
9.32	1.93560583165186\\
9.34	1.91695212531351\\
9.36	1.90955107708862\\
9.38	1.89086317319521\\
9.42	1.84597270861478\\
9.44	1.83393200666023\\
9.5	1.83268056601669\\
9.86	1.83518344730377\\
9.88	1.84597270861478\\
9.9	1.83518344730377\\
9.94	1.82466192375037\\
9.98	1.80660108480079\\
10	1.80660108480079\\
};
\addplot [color=mycolor5, forget plot, line width = 0.6pt]
  table[row sep=crcr]{%
0	29.6140448603194\\
0.0800000000000001	30.4187719910119\\
0.119999999999999	32.1542924768201\\
0.16	30.4187719910119\\
0.199999999999999	30.0164084256657\\
0.24	30.0164084256657\\
0.279999999999999	29.6140448603194\\
0.32	29.6140448603194\\
0.359999999999999	29.1347089627201\\
0.4	28.2807216500064\\
0.44	27.8013857524071\\
0.48	26.9473984396935\\
0.52	25.204207105772\\
0.540000000000001	24.3326114388113\\
0.58	24.2915490854421\\
0.640000000000001	24.2915490854421\\
0.66	25.1631447524028\\
0.68	25.204207105772\\
0.720000000000001	26.9063360863243\\
0.74	27.059570426919\\
0.76	27.4733378477498\\
0.779999999999999	27.7881595168811\\
0.82	28.3426491348121\\
0.84	28.6311084351746\\
0.859999999999999	28.9961746067259\\
1.14	28.9961746067259\\
1.16	29.1613171589227\\
1.18	29.8047102057813\\
1.2	30.3131867809103\\
1.22	30.7304123976162\\
1.24	31.4052869726007\\
1.26	32.198132151255\\
1.28	32.5645679351689\\
1.56	32.5645679351689\\
1.58	32.8954822081384\\
1.6	33.5517222567788\\
1.62	34.0454672162131\\
1.64	34.2324781287866\\
1.84	34.6172996665456\\
2.04	34.8926373967466\\
2.32	35.029240445457\\
2.42	35.0699541188159\\
3.32	35.0188984401705\\
3.36	34.9059536724701\\
3.38	34.735615773559\\
3.4	34.4919953016368\\
3.42	33.7411329187772\\
3.44	32.0137448666267\\
3.46	30.1167756943869\\
3.48	28.99440505188\\
3.5	28.3749921371366\\
3.52	27.2976151110236\\
3.56	23.4990228233172\\
3.58	22.3244161618441\\
3.6	21.3212216583498\\
3.62	20.0811484826561\\
3.66	16.7089019473367\\
3.68	15.3540215022201\\
3.72	12.7762814939424\\
3.74	11.4369018769304\\
3.76	9.72320596729002\\
3.78	7.85126376201405\\
3.8	6.96138847714383\\
3.82	6.90603323620355\\
3.84	6.77156393018844\\
3.86	6.59063484118045\\
3.9	6.31374790544389\\
3.92	6.1955673884329\\
3.94	6.14870481669041\\
3.96	6.14804871374249\\
3.98	6.063646670353\\
4	5.88310849543003\\
4.02	5.66107404312346\\
4.04	5.41849580996335\\
4.06	5.27658031781502\\
4.08	5.25200084100164\\
4.26	5.25200084100164\\
4.3	5.21851061909514\\
4.32	5.15505258160425\\
4.34	5.07725309723762\\
4.36	5.06130920764353\\
4.56	5.06130920764353\\
4.58	5.03371216818053\\
4.6	4.9135149715082\\
4.62	4.81002688892523\\
4.64	4.78321838994461\\
4.7	4.67053619563189\\
4.72	4.65487359861579\\
4.74	4.61784521297435\\
4.76	4.54905577793674\\
4.78	4.43521326087778\\
4.8	4.29745193201145\\
4.84	4.11512624340223\\
4.86	4.05344324570845\\
5.02	4.05344324570845\\
5.04	3.99096364398782\\
5.08	3.75741313146293\\
5.14	3.34415422999731\\
5.18	2.94251298538828\\
5.2	2.77892162355859\\
5.22	2.65335684740794\\
5.24	2.59629107196597\\
5.26	2.5598909708467\\
5.3	2.47145526105686\\
5.32	2.38139210484033\\
5.34	2.2528820256705\\
5.36	2.1611198749856\\
5.38	2.13387523365392\\
5.44	2.13387523365392\\
5.46	2.12836615526455\\
5.48	2.11802061559222\\
5.5	2.06254171153215\\
5.52	1.96260945397718\\
5.54	1.89783624382641\\
5.58	1.81477709784002\\
5.6	1.78707876537808\\
5.62	1.77013650350744\\
5.64	1.75936119000158\\
5.66	1.72457216777665\\
5.68	1.6323346771786\\
5.7	1.53145502171289\\
6.2	1.532560907699\\
6.22	1.65336451716774\\
6.24	1.81562728337344\\
6.26	1.8975392519617\\
6.28	1.93021505201726\\
6.3	1.96955509076525\\
6.32	1.98072560855241\\
6.34	2.01176384064173\\
6.36	2.08612488007304\\
6.38	2.18028084159126\\
6.4	2.25253882841592\\
6.42	2.30696067822928\\
6.44	2.37551717090143\\
6.46	2.42652154294616\\
6.48	2.43746016038892\\
7.14	2.43467896082131\\
7.16	2.41519787184526\\
7.2	2.34805851642125\\
7.22	2.31703258510686\\
7.24	2.29864739331769\\
7.32	2.2962638384757\\
7.34	2.27637536335683\\
7.36	2.24971381999374\\
7.4	2.21490826477264\\
7.44	2.20054311137459\\
7.48	2.1721701515416\\
7.5	2.16749289999718\\
7.54	2.16749289999718\\
7.56	2.15987725432213\\
7.6	2.11899867605334\\
7.62	2.11141316124979\\
8.2	2.11141316124979\\
8.22	2.12576572170658\\
8.24	2.15930130570524\\
8.26	2.17848432924713\\
8.36	2.17848432924713\\
8.38	2.18693284207242\\
8.4	2.21440448831895\\
8.42	2.249146665326\\
8.46	2.2869536039294\\
8.48	2.31545122384359\\
8.5	2.3287818035282\\
8.6	2.33070732376612\\
8.7	2.3580102485846\\
8.78	2.3580102485846\\
8.8	2.36617364500444\\
8.84	2.42777014457116\\
8.86	2.44332022161468\\
8.9	2.45825194604674\\
8.92	2.47517594288929\\
8.94	2.50408104613165\\
8.96	2.52246915327365\\
8.98	2.52702367585485\\
9.04	2.55793832747488\\
9.06	2.56442002536921\\
9.22	2.56442002536921\\
9.24	2.59643762494127\\
9.26	2.68788547493986\\
9.28	2.79495424071419\\
9.3	2.842592756062\\
9.38	2.842592756062\\
9.4	2.85140085846616\\
9.42	2.87423791751036\\
9.44	2.88826687415039\\
9.62	2.89358076169578\\
9.66	2.91109721711143\\
9.82	2.92207484802541\\
9.86	2.93610373500794\\
10	2.93610373500794\\
};
\addplot [color=mycolor6, forget plot, line width = 0.6pt]
  table[row sep=crcr]{%
0	26.1754903163487\\
0.619999999999999	26.1294642123252\\
0.68	26.05447805678\\
0.98	25.9099238429521\\
1	25.7679022994698\\
1.02	25.102610691752\\
1.04	23.7841205605608\\
1.06	22.2401674384443\\
1.08	21.0909520925309\\
1.1	20.6290852073115\\
1.12	20.4816670743827\\
1.14	20.2483709034982\\
1.16	19.7511161686525\\
1.2	19.1086939731642\\
1.22	18.5339717176354\\
1.24	18.044861185867\\
1.3	17.4161110605978\\
1.4	17.4161110605978\\
1.42	17.1310394918081\\
1.44	16.7141188383301\\
1.48	16.1503034591524\\
1.5	15.813168431327\\
1.52	15.6323620950185\\
1.6	15.6323620950185\\
1.64	15.1733106514909\\
1.66	15.1676423836199\\
1.7	14.9172320003708\\
1.72	14.8942083778603\\
1.78	14.6669848086173\\
1.8	14.5186357690401\\
1.82	14.4397612393744\\
1.86	14.1748191145896\\
1.88	14.1032061041378\\
1.9	13.9618898004948\\
1.94	13.7697561401775\\
1.98	13.0319431078048\\
2	12.8398624772921\\
2.02	12.2927363260099\\
2.04	11.2297439108176\\
2.06	10.3229379676026\\
2.08	9.98635719642347\\
2.1	9.83222577508957\\
2.12	9.34835870126287\\
2.14	8.4896104754223\\
2.16	8.02128237286424\\
2.2	8.02128237286424\\
2.22	7.77719849272682\\
2.24	7.36484295701512\\
2.26	6.72078089791663\\
2.28	5.81299019498502\\
2.3	5.31573635291283\\
2.32	5.06227570648545\\
2.34	4.71613344342393\\
2.36	4.48650047827034\\
2.38	4.30066208724965\\
2.4	4.05122102543663\\
2.42	3.83895148024222\\
2.44	3.76096183000734\\
2.46	3.72958901872249\\
2.5	3.57422934678145\\
2.52	3.5490971847952\\
2.56	3.54671589691662\\
2.58	3.53084819552804\\
2.6	3.46917920540437\\
2.64	3.30476853904606\\
2.68	3.10911557041121\\
2.7	3.00240894995376\\
2.72	2.93417834962876\\
2.76	2.89960590730616\\
2.78	2.88741579746541\\
2.8	2.86355359771547\\
2.82	2.81502668058656\\
2.84	2.74323777975874\\
2.86	2.63478692717018\\
2.88	2.54543933031289\\
2.9	2.48621079455752\\
2.92	2.45125990486106\\
2.96	2.45125990486106\\
3	2.38610357383896\\
3.04	2.38610357383896\\
3.06	2.36549625075806\\
3.08	2.32371075133679\\
3.1	2.30253257499643\\
3.2	2.30253257499643\\
3.22	2.28540513636546\\
3.24	2.23183891494153\\
3.26	2.21546910889264\\
3.74	2.21546910889264\\
3.76	2.28728066874725\\
3.78	2.35017338575336\\
3.8	2.37007100586577\\
3.82	2.40675592466736\\
3.86	2.54233809647938\\
3.9	2.65427829350954\\
3.92	2.6871393752194\\
3.96	2.6871393752194\\
3.98	2.71751133500975\\
4	2.76474052211082\\
4.02	2.79594716501496\\
4.04	2.81209464933823\\
4.06	2.86009859439201\\
4.08	2.96018454072556\\
4.1	3.02567435532793\\
4.16	3.12243328051832\\
4.22	3.17760419438293\\
4.24	3.19784055525473\\
4.28	3.25810284848869\\
4.3	3.27434814560501\\
4.78	3.27434814560501\\
4.8	3.24139714322964\\
4.82	3.12429550786534\\
4.84	2.93829418903102\\
4.86	2.75135580091651\\
4.88	2.64223762553416\\
4.9	2.56512875681411\\
4.92	2.47703840408015\\
4.96	2.32370669293978\\
4.98	2.26864727701137\\
5	2.19949307989793\\
5.02	2.0027034236759\\
5.04	1.68160834334829\\
5.06	1.38583752817635\\
5.08	1.21417586415329\\
5.1	1.15840121904103\\
5.64	1.16019175834899\\
5.66	1.18302394688121\\
5.68	1.21223513659282\\
5.7	1.22603290672544\\
5.72	1.26038658891937\\
5.76	1.41676431873434\\
5.78	1.47189084875039\\
5.82	1.55423957388692\\
5.84	1.58582840911574\\
5.88	1.61840871181077\\
5.9	1.6507668388376\\
5.92	1.69315923918697\\
5.94	1.73164838005039\\
5.96	1.76177122373633\\
5.98	1.78206486566875\\
6	1.81202706027263\\
6.02	1.86868616297134\\
6.04	1.93180247721409\\
6.08	2.0249201022762\\
6.1	2.09026346051048\\
6.12	2.21893722922309\\
6.14	2.39585149761527\\
6.16	2.49762440594799\\
6.18	2.51904295404301\\
6.26	2.57729099946538\\
6.28	2.60334812795724\\
6.32	2.66880911235826\\
6.36	2.72914073303424\\
6.38	2.75908325273188\\
6.4	2.77383055931785\\
6.42	2.80644100676437\\
6.44	2.87905012246326\\
6.46	2.99919206959769\\
6.48	3.14980931819248\\
6.5	3.26627263417527\\
6.52	3.32529901989908\\
6.56	3.39494399031\\
6.58	3.41138256240312\\
6.76	3.41138256240312\\
6.78	3.43321362819748\\
6.8	3.4691645701909\\
6.82	3.52125861890804\\
6.84	3.61420036506492\\
6.86	3.75404370278972\\
6.88	3.9473311687545\\
6.9	4.05574287063329\\
6.94	4.05574287063329\\
6.96	4.16238020104402\\
6.98	4.34558322377132\\
7	4.56748408867731\\
7.02	4.86435540864279\\
7.04	5.05768010690926\\
7.1	5.2673184531558\\
7.12	5.30188563809092\\
7.66	5.30188563809092\\
7.68	5.19675728590739\\
7.7	4.94435036366971\\
7.72	4.73411352005872\\
7.74	4.60035509584953\\
7.76	4.42241841322701\\
7.78	4.23167533105328\\
7.8	4.09145708625226\\
7.82	4.00781532331345\\
7.88	3.88308119003159\\
7.92	3.87565235970214\\
8	3.87565235970214\\
8.04	3.80346008183383\\
8.12	3.80103053513633\\
8.16	3.75078642608479\\
8.36	3.73297881908743\\
8.38	3.73297881908743\\
8.4	3.72063963754525\\
8.42	3.69787500456885\\
8.48	3.66110276990904\\
8.5	3.64654866027033\\
8.58	3.64654866027033\\
8.6	3.62126752772468\\
8.62	3.56055534169605\\
8.64	3.52512428821307\\
8.7	3.52512428821307\\
8.72	3.46274508552468\\
8.74	3.33060294537954\\
8.76	3.26084000792277\\
9.02	3.26084000792277\\
9.04	3.21941519466875\\
9.52	3.21941519466875\\
9.54	3.17799038141474\\
9.56	3.18483984743492\\
9.58	3.28424766981208\\
9.62	3.54980939128064\\
9.7	3.62712414295811\\
9.74	3.62712414295811\\
9.78	3.70075732294895\\
9.86	3.70075732294895\\
9.9	3.75040298586655\\
9.94	3.8538563959167\\
9.98	3.88675518069465\\
10	3.89547890526118\\
};
\addplot [color=mycolor7, forget plot, line width = 0.6pt]
  table[row sep=crcr]{%
0	6.58012632094373\\
0.0800000000000001	6.58012632094373\\
0.119999999999999	6.61921822236907\\
0.76	6.61921822236907\\
0.779999999999999	6.63876417308175\\
0.800000000000001	8.04617534968483\\
0.82	9.25764831840708\\
1.3	9.25764831840708\\
1.32	8.71735033327779\\
1.34	8.12371368251457\\
1.36	7.67054203032769\\
1.38	7.4471013009429\\
1.44	7.4471013009429\\
1.46	7.04631045524819\\
1.5	5.54660759035134\\
1.52	4.75780085702003\\
1.54	3.95040502918962\\
1.56	3.56152082701225\\
1.58	3.47204440860118\\
1.6	3.42863455665546\\
1.62	3.42863455665546\\
1.64	3.38798988029896\\
1.66	3.32468214809349\\
1.68	3.24298701752756\\
1.72	2.93674258927738\\
1.74	2.84565015156111\\
1.76	2.77035955704011\\
1.78	2.72480694123544\\
1.8	2.70190205260429\\
1.84	2.67263496755414\\
1.86	2.63556987258677\\
1.9	2.54266496477641\\
1.92	2.49749240611317\\
1.94	2.44556884264662\\
1.96	2.42165154234674\\
2.04	2.42066365792158\\
2.06	2.40285027745357\\
2.08	2.37603850277838\\
2.16	2.3407472711467\\
2.36	2.33603461637358\\
2.5	2.33156171696678\\
2.78	2.33155950691835\\
2.82	2.29559558461415\\
2.84	2.29559337456571\\
2.88	2.2347037034521\\
2.94	2.23415177253296\\
2.98	2.2051913998588\\
3	2.17187971120576\\
3.02	2.10448841732849\\
3.04	2.06153166272077\\
3.06	2.05024037834903\\
3.1	1.97580787157522\\
3.14	1.97580787157522\\
3.18	2.04635587927537\\
3.2	2.05764716364711\\
3.22	2.10448841732849\\
3.26	2.17335007529117\\
3.28	2.22068273369815\\
3.3	2.2347037034521\\
3.32	2.2347037034521\\
3.34	2.25140353471453\\
3.36	2.28184837027133\\
3.4	2.31357644074203\\
3.44	2.34594096895788\\
3.46	2.36032243099743\\
3.48	2.36147849340117\\
3.52	2.41830534568761\\
3.58	2.42031701146963\\
3.6	2.42031701146963\\
3.64	2.36032243099743\\
3.66	2.33643657565358\\
3.76	2.33643657565358\\
3.78	2.39126196786946\\
3.8	2.4222015047415\\
3.82	2.44510008369415\\
3.84	2.4764947023869\\
3.86	2.484990742127\\
3.98	2.484990742127\\
4	2.50266427433568\\
4.02	2.53675815825854\\
4.04	2.55416509177698\\
4.08	2.60160208892449\\
4.1	2.61374902110741\\
4.12	2.65843608377865\\
4.14	2.69126519089492\\
4.16	2.69126519089492\\
4.18	2.7019276157934\\
4.2	2.82427233785777\\
4.22	3.02125103759216\\
4.24	3.12141289421955\\
4.28	3.1846308814251\\
4.3	3.2134225158862\\
4.32	3.29397457697093\\
4.34	3.45977412341747\\
4.36	3.6107903473667\\
4.38	3.73705378387002\\
4.4	3.81407462953508\\
4.54	3.81800304889457\\
4.58	3.97111758918066\\
4.62	4.0094515466305\\
4.74	4.01927775133766\\
5.08	4.01237555881938\\
5.12	3.96637994103776\\
5.16	3.81800304889457\\
5.18	3.80685258001873\\
5.2	3.71734643635051\\
5.22	3.61795817523771\\
5.24	3.60035869706806\\
5.32	3.59986338585932\\
5.34	3.58544561386312\\
5.36	3.53083673666286\\
5.38	3.49064563145878\\
5.4	3.49064563145878\\
5.42	3.45954374204493\\
5.46	3.34663568538148\\
5.48	3.2671454770225\\
5.52	3.07218137282538\\
5.54	3.00569065291891\\
5.56	2.91993248996239\\
5.58	2.80360875578588\\
5.6	2.73212699980289\\
5.62	2.69726734619923\\
5.64	2.61239700045083\\
5.66	2.47557202203264\\
5.68	2.37341373883909\\
5.7	2.34277639815406\\
5.96	2.34277639815406\\
5.98	2.31379389733286\\
6	2.25731392616933\\
6.02	2.229816455827\\
6.36	2.229816455827\\
6.38	2.22051465059696\\
6.4	2.15395770772715\\
6.44	1.92606518719721\\
6.46	1.85755379954136\\
6.48	1.82610924404816\\
6.5	1.81567216557219\\
6.56	1.81427164195027\\
6.76	1.81427164195027\\
6.78	1.80343638891741\\
6.8	1.7787453885768\\
6.82	1.75052708070609\\
6.84	1.69653922999968\\
6.88	1.51647612231593\\
6.9	1.4829968124587\\
6.92	1.47783515612202\\
6.94	1.42742926179243\\
6.96	1.32520645873579\\
6.98	1.25619199143224\\
7	1.20178416780717\\
7.02	1.12797192278041\\
7.04	1.08137146059071\\
7.06	1.06757654911865\\
7.08	1.06486435560231\\
7.12	1.06368691436461\\
7.16	1.06250947312691\\
7.18	1.04131374380927\\
7.2	0.989033967358837\\
7.22	0.947831662872055\\
7.24	0.932598893484534\\
7.26	0.927484381450999\\
7.74	0.927484381450999\\
7.76	0.963682940617337\\
7.78	1.03878354181563\\
7.8	1.0943019322999\\
7.82	1.1201608704555\\
7.84	1.13896544028315\\
7.86	1.17395272841839\\
7.88	1.22223679197674\\
7.9	1.25416851256955\\
7.96	1.31117603551673\\
7.98	1.31944500653356\\
8	1.33396568309293\\
8.02	1.3716257716292\\
8.04	1.44430507809327\\
8.06	1.5325989178171\\
8.08	1.58561706190324\\
8.1	1.60090080735628\\
8.14	1.61999525571437\\
8.16	1.63261660947072\\
8.18	1.65451216556768\\
8.2	1.68603532999394\\
8.22	1.73029022628207\\
8.24	1.80236953653953\\
8.26	1.85251032245683\\
8.28	1.85747233116714\\
8.34	1.85747233116714\\
8.38	1.86786560161049\\
8.4	1.86786560161049\\
8.46	1.88052900210858\\
8.54	1.89319240260668\\
8.64	1.8959434936942\\
8.68	1.90211968954056\\
8.8	1.90402283169142\\
8.84	1.91092785363683\\
8.92	1.91359153600413\\
8.96	1.93051270784238\\
9.2	1.93051270784238\\
9.22	1.94248320731033\\
9.24	1.97199696891064\\
9.26	2.01299647377204\\
9.28	2.04108953669882\\
9.34	2.07140398103657\\
9.38	2.10708438353761\\
9.42	2.1968671979609\\
9.58	2.2004619598242\\
9.64	2.21332908066372\\
9.66	2.21913071001828\\
10	2.21913071001828\\
};
\end{axis}
\end{tikzpicture}% 

%% file: img/experiments/comparison.tex
\definecolor{mycolor1}{rgb}{0.74902,0.00000,0.00000}%
\definecolor{mycolor2}{rgb}{0.00000,0.74902,0.00000}%
\definecolor{mycolor3}{rgb}{0.00000,0.00000,0.74902}%
\begin{tikzpicture}

\begin{axis}[%
width=\figW,
height=0.944\figH,
at={(0\figW,0\figH)},
scale only axis,
xmin=0,
xmax=211.64,
xlabel style={font=\color{white!15!black}},
xlabel={Time $t$ $[\mathrm{s}]$},
ymin=0,
ymax=40,
y label style={at={(axis description cs:0.1,.5)},anchor=south},
ylabel={Error $\epsilon\,[\g]$},
axis background/.style={fill=white},
xmajorgrids,
ymajorgrids,
grid style={dotted},
legend style={at={(0.5,1.03)}, anchor=south, legend columns=3, legend cell align=left, align=left, draw=white!15!black}
]
\addplot [color=mycolor1, line width=1.0pt]
  table[row sep=crcr]{%
0	13.168133376499\\
0.439999999999998	37.0102019613421\\
0.879999999999995	31.9503645210216\\
1.31999999999999	33.2787689660221\\
1.75999999999999	35.0060372666964\\
2.19999999999999	35.3183759142315\\
2.63999999999999	35.3284884503239\\
3.08000000000001	35.1701026774685\\
3.52000000000001	27.196648514087\\
3.96000000000001	22.255189410989\\
4.40000000000001	22.0885149676914\\
4.84	4.3346120954547\\
5.28	2.63686973732905\\
5.72	1.47383371229387\\
6.16	1.45646518723206\\
6.59999999999999	2.26520106387539\\
7.03999999999999	2.308829306039\\
7.47999999999999	1.91519885196422\\
7.91999999999999	1.72405290589799\\
8.36000000000001	2.01714989975548\\
8.80000000000001	2.04082783284511\\
9.24000000000001	2.45890859006033\\
9.68000000000001	2.76339180166048\\
10.12	2.71456735300279\\
10.56	2.73016704216977\\
11	2.62005271186544\\
11.44	1.44767996678442\\
11.88	0.765563476667637\\
12.32	0.584361351174266\\
12.76	0.849085926864092\\
13.2	1.66397892712951\\
13.64	2.35754917689763\\
14.08	2.65270158707369\\
14.52	2.31877588141614\\
14.96	2.29644768275213\\
15.4	2.07491788473064\\
15.84	1.91457546054815\\
16.28	1.9092240692508\\
16.72	1.77713438080193\\
17.16	1.62156019349024\\
18.04	1.61525425567086\\
18.48	1.76846192491382\\
18.92	1.80386590420616\\
19.36	1.81246508135902\\
19.8	1.73366271179771\\
20.24	1.48442340544145\\
20.68	1.5134306857781\\
21.12	1.86876751472411\\
21.56	2.01258840998227\\
22	2.01258840998227\\
22.44	1.96043004558689\\
22.88	1.96292992794091\\
23.32	1.76160311791173\\
23.76	1.49433799470253\\
24.2	1.60555937354931\\
24.64	1.97540738729671\\
25.08	2.06565229315737\\
25.52	2.4697135197259\\
25.96	2.7626793670911\\
26.4	2.84112503456487\\
26.84	2.22075832583127\\
27.28	1.01743866470449\\
27.72	0.775993629589181\\
28.16	1.20065357420262\\
28.6	1.65382281670847\\
29.04	2.36764237106829\\
29.48	2.85540545562978\\
29.92	3.3102294332453\\
30.36	2.47206167327266\\
30.8	2.47206167327266\\
31.24	3.37786152229177\\
31.68	2.60748939756891\\
32.12	2.59129722749114\\
32.56	3.67024394912386\\
33	3.33288605122866\\
33.44	3.33288605122866\\
33.88	3.90690181652562\\
34.32	3.8370063961585\\
34.76	3.94229743728636\\
35.2	4.13249778228047\\
35.64	4.58480023517177\\
36.08	4.34967067411068\\
36.52	3.91668944555497\\
36.96	3.67465883357525\\
37.4	4.39483344865158\\
37.84	4.83197418986458\\
38.28	4.98793861644918\\
38.72	5.26433521864266\\
39.16	5.30785983336955\\
39.6	4.7262204795295\\
40.04	4.56607814362127\\
40.92	4.94224222146917\\
41.36	4.94134742753334\\
42.24	4.8828232902122\\
42.68	4.81819898857401\\
43.56	4.53866712085713\\
44	4.47567794122853\\
44.44	4.56459601119133\\
44.88	4.78151880532323\\
45.32	4.82175372637747\\
45.76	4.77741109902126\\
46.2	4.28329176568812\\
46.64	4.27276677264126\\
47.08	4.48094868228071\\
47.52	5.0963211275093\\
47.96	5.04769878001483\\
48.4	4.68910086212347\\
48.84	4.6505811859264\\
49.28	4.67071191019863\\
49.72	5.46708283893699\\
50.16	6.16633994491295\\
50.6	6.30547640172929\\
51.04	5.89985884897268\\
51.48	5.61589279313483\\
51.92	5.21858218066188\\
52.8	5.11549741558068\\
53.24	4.64474404350759\\
53.68	4.45047363045546\\
54.12	4.44621927935279\\
54.56	4.57552020050366\\
55	5.47614992570243\\
55.44	5.46306503383522\\
55.88	4.84554862147371\\
56.32	4.29624693611279\\
56.76	4.34518068244367\\
57.2	4.64635732841498\\
57.64	5.11902189043695\\
58.08	5.20529226159644\\
58.52	5.13840844915021\\
58.96	5.12170702262748\\
59.4	5.13624529005352\\
59.84	5.08604915534139\\
60.28	4.89965226841434\\
60.72	4.73264538713758\\
61.16	4.82379576375263\\
61.6	5.2270629116461\\
62.04	5.15525813804476\\
62.48	4.92668258922686\\
62.92	4.66755253015464\\
63.36	4.62311961747236\\
63.8	4.66102363871704\\
64.24	4.75453871558565\\
65.12	4.68034713945491\\
65.56	4.88629541779699\\
66	5.07343740805419\\
66.88	5.35883610611558\\
67.76	4.97054237008504\\
68.2	4.99065660439902\\
68.64	4.65931107714971\\
69.08	4.54818511948375\\
69.52	4.6962402801299\\
69.96	4.44067105480309\\
70.4	3.58718850214461\\
70.84	3.12720125484708\\
71.28	2.56151559767085\\
72.16	2.31602443517517\\
72.6	1.86002153157872\\
73.04	1.84811758292372\\
73.48	1.95965419453188\\
73.92	2.40540430719892\\
74.36	2.09124227549177\\
74.8	1.49208190545087\\
75.24	1.45951139082746\\
75.68	2.20211033554415\\
76.12	2.50706081041423\\
76.56	3.45749415140625\\
77	3.30835005307802\\
77.44	1.92229252231215\\
77.88	1.63113248479328\\
78.32	2.15459903511223\\
78.76	3.51026475692785\\
79.2	2.69068733718208\\
79.64	3.33232765344798\\
80.08	3.55882337425084\\
80.52	2.86223449237326\\
80.96	3.05557289292543\\
81.4	3.01644018962162\\
81.84	1.31607815339666\\
82.28	1.13696227448028\\
82.72	1.36040658453308\\
83.16	2.06042280781429\\
83.6	2.43673482272226\\
84.04	2.4922224270826\\
84.48	1.66548337292249\\
84.92	1.08748399081287\\
85.36	1.363032314767\\
85.8	2.20618719795428\\
86.24	3.08899775103967\\
86.68	2.63942852865478\\
87.12	2.66052720154295\\
87.56	2.80394172265292\\
88	2.17580042731211\\
88.44	2.15281095532322\\
88.88	2.17814868912257\\
89.32	1.68749094064989\\
89.76	1.54384083582505\\
90.2	1.24111603767261\\
90.64	1.80806086647763\\
91.08	2.07757707149278\\
91.52	2.10776913011608\\
91.96	2.34553269428542\\
92.4	2.19409609674176\\
92.84	1.99453629215085\\
93.28	2.17682761039995\\
93.72	2.12294271027787\\
94.16	1.78990240789801\\
94.6	1.88291870075685\\
95.04	1.85390863460941\\
95.48	1.07464101615213\\
95.92	1.07648781816022\\
96.8	2.25342027128181\\
97.24	2.21181788927308\\
97.68	2.09469893854063\\
98.12	1.86821386749114\\
98.56	1.56465376752161\\
99	2.30526304039634\\
99.44	2.86940493882292\\
99.88	1.76537321072101\\
100.32	1.83603933475021\\
100.76	2.28126992465263\\
101.2	2.34377545352322\\
101.64	1.82852471365786\\
102.08	1.54151392295381\\
102.52	1.4941317221369\\
102.96	1.4755723323575\\
103.4	1.5566657366976\\
103.84	1.61272221710539\\
104.28	1.86211594666969\\
104.72	2.08750061646086\\
105.16	2.10170985076334\\
105.6	1.97921069268378\\
106.04	2.01625582955413\\
106.48	1.83239074348239\\
106.92	1.77986356355601\\
107.36	1.76116550386149\\
107.8	1.81750909735612\\
108.24	1.81993335875936\\
108.68	1.84933152260695\\
110	1.61730930275257\\
110.44	1.62426233830408\\
110.88	1.66049000685109\\
111.32	1.52319966440595\\
111.76	1.54121943878641\\
112.64	1.75451633405791\\
113.08	1.75886567273582\\
113.52	1.78478932320473\\
113.96	1.89728037292767\\
114.4	1.90777599051637\\
114.84	1.89040489563249\\
115.28	1.96568941683833\\
115.72	1.9548382488563\\
116.6	1.7731860006482\\
117.48	1.90509394757092\\
118.36	1.90329458078881\\
118.8	1.86883284865726\\
119.24	1.92804558059254\\
119.68	1.86490886677109\\
120.12	1.88284139383902\\
120.56	2.01743201893902\\
121	1.91854538904053\\
121.44	2.16287087423964\\
121.88	2.26909164241388\\
122.32	2.25258476743167\\
122.76	2.25739282684347\\
123.64	2.0823338428049\\
124.08	2.08396130129117\\
124.52	1.99435089359949\\
125.4	1.75711368255952\\
125.84	1.73151582750802\\
126.28	1.90887762397421\\
126.72	2.15158311702638\\
127.16	2.13340123690728\\
127.6	1.99296260901986\\
128.04	1.53260510950469\\
128.48	1.13602754412219\\
128.92	1.1303132938836\\
129.36	1.47049515591019\\
129.8	1.76324045148641\\
130.24	2.00571530083539\\
130.68	2.284226441708\\
131.12	2.29935075632281\\
132	2.39461090072757\\
133.32	2.21014148581119\\
133.76	2.17288845219912\\
134.2	2.23748619919846\\
134.64	2.21562196675109\\
135.08	2.27544747340636\\
135.52	2.58765714574747\\
135.96	2.37611754119789\\
136.4	2.10130988402119\\
136.84	2.18679261877585\\
137.28	3.7766212557313\\
137.72	3.53723227466858\\
138.16	3.69763849070449\\
138.6	4.05688184616866\\
139.04	3.92690566845138\\
139.48	3.52941681580549\\
139.92	3.87556536070264\\
140.36	3.8757874647246\\
140.8	3.54396774981635\\
141.24	3.40169554142511\\
141.68	3.97636993090796\\
142.12	3.96593515149226\\
142.56	3.48922600851458\\
143	3.50323725754276\\
143.44	3.85788685226811\\
143.88	3.87967766799636\\
144.32	3.47562137114335\\
144.76	2.1672433324224\\
145.2	2.15258847650375\\
145.64	2.23631177213571\\
146.08	2.04172599171793\\
146.52	1.67453782512573\\
146.96	1.54741312579395\\
147.4	2.20105562975257\\
147.84	2.45694312469973\\
148.28	2.83678131975512\\
148.72	3.99315421841987\\
149.16	4.02982719792644\\
149.6	3.47193175054059\\
150.04	2.47551159699967\\
150.48	2.57784655702204\\
150.92	2.58596342656779\\
151.36	1.84016708615118\\
151.8	1.3276842015112\\
152.24	2.34822201696994\\
152.68	2.78801605583553\\
153.12	2.42543457158871\\
153.56	1.50719028031068\\
154	1.31249073916044\\
154.44	1.73208458560114\\
154.88	2.26603014186608\\
155.32	2.65416041873178\\
155.76	2.97884478902716\\
156.2	2.42996556649942\\
156.64	2.03505866434227\\
157.08	2.04814950587649\\
157.52	1.70121324165311\\
157.96	1.5637109103929\\
158.4	1.85436236029039\\
158.84	2.21972932977442\\
159.28	1.89231241535995\\
159.72	1.15830105705646\\
160.16	1.27240838037369\\
160.6	1.95167376436854\\
161.04	2.43105451352926\\
161.48	2.59055437276245\\
161.92	2.09360037695768\\
162.36	1.80372284165705\\
162.8	1.72718372611163\\
163.24	1.5920437138812\\
164.12	1.71892171645038\\
164.56	1.68006803015007\\
165	1.60856615280434\\
165.44	1.58581409906543\\
166.76	1.57311032656295\\
167.2	1.55578863546975\\
167.64	1.5170924676502\\
168.08	1.49398337301525\\
170.72	1.50011283398905\\
171.16	1.47302786690906\\
172.04	1.48391755562687\\
172.92	1.49364418893919\\
173.36	1.49652504228993\\
174.24	1.48625722174364\\
174.68	1.51680875260641\\
175.56	1.52265298653225\\
176.44	1.51991460263525\\
177.76	1.54553708277658\\
178.2	1.5784160614125\\
179.96	1.62805410527426\\
180.84	1.67538433004884\\
182.16	1.67903352014423\\
182.6	1.63934248843063\\
183.04	1.61839410022623\\
183.48	1.64741435080961\\
184.8	1.64897614496238\\
185.24	1.65905448816113\\
185.68	1.62784264161422\\
186.12	1.53942363551917\\
186.56	1.25473269991818\\
187	1.27368047024558\\
187.44	1.35622590412643\\
187.88	1.77773211456153\\
188.32	1.78513487007805\\
188.76	1.75938994472901\\
189.2	1.68496285678194\\
189.64	1.43277222580224\\
190.08	1.39599989260037\\
190.52	1.46312266585272\\
190.96	1.47697695089718\\
191.4	1.46937118448466\\
191.84	1.42703154685236\\
192.72	1.4682652383043\\
193.16	1.5537876339678\\
193.6	1.60957087994248\\
194.04	1.57962358452758\\
194.48	1.36784180795053\\
194.92	1.48519163556705\\
195.36	1.78142730854631\\
195.8	1.78263784350216\\
196.24	1.63863328129281\\
196.68	1.63863328129281\\
197.12	1.42027453502098\\
197.56	1.4544248915283\\
198	1.80668895062979\\
198.44	1.7321705673624\\
198.88	1.68922944290395\\
199.32	1.80073116574303\\
199.76	1.71993780189308\\
200.2	1.74653871204842\\
200.64	1.98834432536074\\
201.08	2.0918898410244\\
201.96	2.0123656719239\\
202.4	1.90646107284462\\
202.84	1.51783951954889\\
203.28	1.4947665095392\\
203.72	1.92975975240878\\
204.16	1.92975975240878\\
204.6	1.64736045170648\\
205.04	1.65731551146973\\
205.48	1.57143109664096\\
205.92	1.26714888983165\\
206.36	1.40791259675757\\
207.24	1.54158552707622\\
208.12	1.70879906274695\\
208.56	2.0287826408217\\
209	2.14957305908152\\
209.44	2.20947353840577\\
209.88	2.29049061833834\\
210.32	2.26807648437045\\
211.2	2.36963152310324\\
};
\addlegendentry{KC}

\addplot [color=mycolor2, line width=1.0pt]
  table[row sep=crcr]{%
0	1.42629204863499\\
0.879999999999995	1.47508476761806\\
1.31999999999999	1.71239528501772\\
1.75999999999999	1.63155723283742\\
2.19999999999999	1.74077885485457\\
3.08000000000001	1.79214456696036\\
3.52000000000001	2.25088150620689\\
3.96000000000001	2.64756522072048\\
4.40000000000001	2.00914999525952\\
4.84	1.7527752607318\\
5.28	2.04192767721815\\
5.72	2.49036732178635\\
6.16	2.87300611640578\\
6.59999999999999	3.93246029408922\\
7.03999999999999	3.34465431854045\\
7.47999999999999	3.38421810849266\\
7.91999999999999	3.39181209214499\\
8.36000000000001	3.1493174820414\\
8.80000000000001	3.40969601709858\\
9.24000000000001	3.46308362567089\\
9.68000000000001	2.69605530545778\\
10.12	2.27668509900795\\
10.56	1.42347161790045\\
11	2.03950410721421\\
11.44	2.78669032328222\\
11.88	2.87901772985381\\
12.32	2.80153755928978\\
12.76	2.51043064992203\\
13.2	2.52413585908147\\
13.64	2.50656113412398\\
14.08	3.17104457507847\\
14.52	2.28745123588496\\
14.96	2.70080980258504\\
15.4	2.900226298989\\
15.84	2.16061561293793\\
16.28	2.37073265640595\\
16.72	3.55222121671446\\
17.16	3.73425436038505\\
17.6	3.70326667850006\\
18.04	2.37549079419554\\
18.48	2.00628948907362\\
18.92	1.42405791614055\\
19.36	1.3800203234666\\
19.8	1.93665644071135\\
20.24	2.25788747563732\\
20.68	3.07811704458993\\
21.12	4.0178628954819\\
21.56	4.38298893268561\\
22	4.38298893268561\\
22.44	4.32371600581999\\
22.88	4.33894086057143\\
23.32	3.87562487439021\\
23.76	2.8953282670231\\
24.2	2.79838110015493\\
24.64	3.28255980209585\\
25.08	3.94328370474514\\
25.52	4.74692011334503\\
25.96	5.09283134684205\\
26.4	5.16032817118719\\
26.84	4.46036155742485\\
27.28	3.15708460358863\\
27.72	2.38128210345835\\
28.16	2.33375979172257\\
28.6	3.24552539956895\\
29.04	3.76360676443196\\
29.48	5.04281016459194\\
29.92	5.59342492547458\\
30.36	4.61318581045509\\
30.8	4.61318581045509\\
31.24	5.60639433093223\\
31.68	4.70389760144803\\
32.12	4.69033143910383\\
32.56	5.95662918662572\\
33	5.51222325689798\\
33.44	5.51222325689798\\
33.88	6.18447475139226\\
34.32	6.122405046757\\
34.76	6.19082042549761\\
35.2	6.33856428959317\\
35.64	6.86185735614006\\
36.08	6.63946990204565\\
36.52	6.18923747741439\\
36.96	5.92622046467673\\
37.4	6.64383226155618\\
37.84	7.14594196073116\\
38.28	7.34839753439826\\
38.72	7.64182827641119\\
39.16	7.780591356477\\
39.6	7.80938022539402\\
40.04	7.71919736956411\\
40.48	7.72208025135018\\
40.92	8.01606856761609\\
41.36	8.06398854327088\\
42.24	8.12637954729081\\
42.68	8.12079834396161\\
43.12	8.05605456757917\\
43.56	7.95533568858912\\
44	7.91407057780447\\
44.44	8.02768691805696\\
44.88	8.27080117944828\\
45.32	8.4676098214552\\
45.76	8.47666627048213\\
46.2	8.16298152233682\\
46.64	8.14698521311126\\
47.08	8.43475270221427\\
47.52	9.16162547215021\\
47.96	9.13059802499356\\
48.4	8.82801257291203\\
48.84	8.82657531590925\\
49.28	8.88699216727696\\
49.72	10.1182034538138\\
50.16	10.5949410187973\\
50.6	10.7564798368946\\
51.04	10.4622636523132\\
51.48	10.2502114260141\\
51.92	9.88450887686938\\
52.36	9.81705406508291\\
52.8	9.80163705850453\\
53.24	9.37770795777448\\
53.68	9.25991879912615\\
54.12	9.34605026646693\\
54.56	10.0981899546465\\
55	10.6134117825941\\
55.44	10.6176257352338\\
55.88	10.0842525073377\\
56.32	9.52111052695295\\
56.76	9.58713545338759\\
57.2	10.0278139883338\\
57.64	10.5699001590497\\
58.08	10.7924230004168\\
58.52	10.7844656236262\\
58.96	10.6746042789713\\
59.4	10.5981916967175\\
59.84	10.5752348213202\\
60.28	10.5767910272152\\
60.72	10.6632052796239\\
61.16	11.1226235060244\\
61.6	11.4425000647964\\
62.04	11.4380029728698\\
62.48	11.2394936678248\\
62.92	11.1695331364315\\
63.36	11.195917750065\\
63.8	11.2612287795922\\
64.24	11.4081536843652\\
64.68	11.4297051641542\\
65.12	11.4799165698839\\
65.56	11.7099076650365\\
66	12.1072060653502\\
66.44	12.4405072951452\\
66.88	12.7356557628922\\
67.32	12.80515686521\\
68.2	13.1321266057268\\
68.64	13.1002072912375\\
69.08	13.1448675377432\\
69.52	13.2456136306047\\
69.96	13.2701284139679\\
70.4	13.3240776008439\\
70.84	13.2726964576734\\
71.28	12.8711655944674\\
71.72	12.3866520298981\\
72.16	11.6961001204558\\
72.6	11.3048497717985\\
73.04	11.2362492215213\\
73.48	11.3438830179824\\
73.92	11.6911772005325\\
74.36	11.7411502184407\\
74.8	11.6597519503346\\
75.24	11.6642530926263\\
75.68	11.9604305324627\\
76.12	12.5867238493119\\
76.56	14.4021724913216\\
77	14.4021724913216\\
77.44	13.277529912734\\
77.88	12.3872936181426\\
78.32	12.5600535097603\\
78.76	13.8109366107189\\
79.2	13.6457331665408\\
79.64	14.6479305923521\\
80.08	14.7166514404103\\
80.52	14.3432660426199\\
81.4	14.3754044572439\\
81.84	12.8336301931366\\
82.28	12.5243059238326\\
82.72	13.0943653132721\\
83.16	14.7926153984156\\
83.6	15.2905308193467\\
84.04	15.2694473252255\\
84.48	14.6035250326727\\
84.92	13.1468636289836\\
85.36	13.0542091938511\\
85.8	13.6705154953002\\
86.24	14.8737564737853\\
86.68	15.222836257198\\
87.12	15.8102028370918\\
87.56	15.9369374813565\\
88	15.8815281897687\\
88.44	15.5945522229827\\
88.88	15.5604592351539\\
89.32	15.0860579747012\\
89.76	14.0611606144753\\
90.2	13.8574516923609\\
90.64	14.1146498200905\\
91.08	14.6942686945796\\
91.52	14.8591714441137\\
91.96	14.8312500179377\\
92.4	14.2885100640505\\
92.84	14.4632624352379\\
93.28	16.0934996559058\\
93.72	16.3177737153044\\
94.16	16.4384085013318\\
95.04	16.3515090658039\\
95.48	14.8485368843454\\
95.92	14.2717073238352\\
96.36	14.7087136228467\\
96.8	16.5668750007456\\
97.24	16.6567379890656\\
97.68	16.6599540903717\\
98.12	16.4063754643077\\
98.56	15.3141129704284\\
99	15.0751356477165\\
99.44	15.9770401268998\\
99.88	16.3013300613032\\
100.32	16.8374134023759\\
100.76	16.6064390297128\\
101.2	16.6692927866881\\
101.64	15.995517954186\\
102.08	15.9572152578538\\
102.52	16.8011965117315\\
102.96	17.0527506768917\\
103.4	17.0147745016031\\
103.84	15.6646731301925\\
104.28	15.0479236090015\\
104.72	14.9791652312787\\
105.16	15.4299372678154\\
105.6	15.9280992339023\\
106.04	16.5414458915128\\
106.48	17.2765984710873\\
106.92	17.2508389100039\\
107.36	17.0135643142462\\
107.8	17.0688971046096\\
108.24	17.6680206950179\\
108.68	17.5917149321689\\
109.12	17.6011931648069\\
109.56	18.0946153265318\\
110	18.253869599864\\
110.44	18.3620926976708\\
110.88	18.5461284904657\\
111.32	18.3462643302644\\
111.76	18.3724033417031\\
112.2	18.7909171314861\\
112.64	18.8500586995212\\
113.08	18.6278894162878\\
113.52	18.695310872299\\
113.96	18.9521266203485\\
114.4	18.9973913812207\\
114.84	18.7263038875846\\
115.28	18.7710259441322\\
115.72	19.0331488569042\\
116.16	18.9391992035593\\
116.6	18.9378982613478\\
117.04	18.9920488922364\\
117.48	19.0213096851718\\
117.92	19.1076597340002\\
118.8	19.1426865018346\\
119.24	19.2790202487924\\
119.68	19.2562558291862\\
120.12	19.2722832225622\\
120.56	19.5982774155712\\
121	19.6287467489144\\
121.44	20.2241187217801\\
121.88	20.4867202944167\\
122.32	20.4938569364182\\
122.76	20.5264323561273\\
123.2	20.4979792744285\\
123.64	20.487072316972\\
124.08	20.5609274424448\\
124.52	20.5903637803795\\
124.96	20.7151463327015\\
125.4	20.8163997010332\\
125.84	20.8362085604527\\
126.28	20.7988892050892\\
126.72	20.6794692851424\\
127.6	20.5086783641561\\
128.04	20.2715124582073\\
128.48	20.0737270179407\\
128.92	19.9602921079806\\
129.36	20.0225257392958\\
129.8	20.1813720976398\\
130.24	20.5721238045489\\
130.68	21.349930681468\\
131.12	21.8780098002702\\
131.56	22.0117534431709\\
132	22.1973614456275\\
132.44	22.3204861217849\\
132.88	22.3677704711197\\
133.32	22.3531119286266\\
133.76	22.3928647109886\\
134.2	22.4772415755542\\
134.64	22.4648986471212\\
135.08	22.477006840255\\
135.52	22.7384698953897\\
136.4	22.254685559596\\
136.84	22.3757675372053\\
137.28	24.2443895525656\\
137.72	24.0257458216093\\
138.16	24.212803803721\\
138.6	24.5665979315969\\
139.04	24.4108485950724\\
139.48	24.0184028833088\\
139.92	24.4604586438998\\
140.36	24.4692541815917\\
140.8	24.2009725844727\\
141.24	24.0301846630882\\
141.68	24.7072975260863\\
142.12	24.6983030655466\\
142.56	24.1989864107745\\
143	24.2420074870689\\
143.44	24.6835247315801\\
143.88	24.6635596955904\\
144.32	22.7287351456615\\
144.76	22.606851412597\\
145.2	21.9646935416837\\
145.64	21.0708088197364\\
146.08	21.5580353595211\\
146.52	22.8007902625066\\
146.96	22.8707263726801\\
147.4	22.8410205381167\\
147.84	22.9497726836504\\
148.28	22.4914455802342\\
148.72	22.7228209154194\\
149.16	22.8611202945507\\
149.6	22.5051945656373\\
150.04	21.1971810527382\\
150.48	21.0248470835571\\
150.92	22.3386603949741\\
151.36	22.9252082681514\\
151.8	22.8733303705054\\
152.24	22.9061050033819\\
152.68	23.137464066233\\
153.12	22.5205118836107\\
153.56	21.7102923609926\\
154	21.0410530934965\\
154.44	21.0415386030411\\
154.88	22.5259941314355\\
155.32	23.5342785473165\\
155.76	23.8685164435955\\
156.2	23.8548066386214\\
156.64	23.8781413125052\\
157.08	23.8781413125052\\
157.52	22.7255383702978\\
157.96	22.7255383702978\\
158.4	23.9845210242625\\
158.84	24.3897875508248\\
159.28	24.1048927802162\\
159.72	22.8796598446625\\
160.16	22.5072546813892\\
160.6	23.9038463116825\\
161.04	24.7393175226403\\
161.48	24.724592385818\\
161.92	23.7601760803901\\
162.36	23.7383690393123\\
162.8	24.2540157322078\\
163.24	24.3449523693496\\
164.12	24.4597182700217\\
164.56	24.5062047448874\\
165	24.6032666195504\\
165.44	24.7342885377979\\
165.88	24.8485567279976\\
166.32	24.9968826387127\\
166.76	25.0761441606624\\
167.64	25.2672886354726\\
168.52	25.3809454073096\\
169.84	25.7230314231429\\
170.72	25.8388555051727\\
171.6	26.0321632134545\\
172.04	26.1097054246786\\
172.48	26.1661578464967\\
172.92	26.2380809428254\\
174.24	26.2933749172885\\
174.68	26.4052160677497\\
175.56	26.5044118628707\\
176.88	26.710141777588\\
177.32	26.810494696765\\
177.76	26.8835077330314\\
178.2	26.9762905844798\\
178.64	27.0374635777109\\
179.08	27.0730194293315\\
181.28	27.3723940596638\\
182.16	27.5614891940773\\
182.6	27.6939140895998\\
184.8	28.0703391262059\\
185.24	28.1147110410791\\
185.68	28.0958447632303\\
186.12	28.0381968806123\\
186.56	27.5661697549109\\
187	27.2741718794335\\
187.44	26.7974796975584\\
187.88	27.6139004392108\\
188.32	29.1197401827074\\
188.76	29.2748791291171\\
189.2	29.2704087752526\\
189.64	29.1872707349139\\
190.08	29.2857032816939\\
190.52	29.3291660071211\\
190.96	28.7442351541602\\
191.4	27.5364352815985\\
191.84	27.7109008984664\\
192.28	27.7854522139424\\
192.72	27.8841444923404\\
193.16	27.9643907604666\\
193.6	28.2866430160529\\
194.04	29.6226458548272\\
194.92	29.9453938879139\\
195.36	30.033355488499\\
195.8	29.9318912610069\\
196.24	29.7272050750062\\
196.68	29.7329195503865\\
197.12	28.9863054842195\\
197.56	28.2805347526192\\
198	28.6155062308706\\
198.44	29.9915036216227\\
198.88	30.1842532009219\\
199.32	30.4180686805124\\
199.76	30.363484438905\\
200.2	30.4659794343447\\
200.64	30.6104060548207\\
201.08	30.7758778370157\\
201.52	31.0384581835431\\
201.96	30.724067676733\\
202.4	30.4954203481522\\
202.84	30.3781560136626\\
203.28	28.8495177271004\\
203.72	28.7940830034925\\
204.16	30.4190060475858\\
204.6	31.0388089025775\\
205.04	31.0567702337512\\
205.48	31.0004667952921\\
205.92	30.6293898584555\\
206.36	29.447189920582\\
206.8	28.9291656483892\\
207.24	29.0004478132525\\
207.68	29.901386756173\\
208.12	30.5574605585748\\
208.56	31.1538909666837\\
209.44	31.5773241680931\\
209.88	31.6746793883955\\
210.32	31.6790976290562\\
211.2	31.7734301473453\\
211.64	31.7882169029697\\
};
\addlegendentry{6D}

\addplot [color=mycolor3, line width=1.0pt]
  table[row sep=crcr]{%
0	6.09866992836155\\
0.439999999999998	5.97352509285676\\
0.879999999999995	5.91502063473541\\
1.31999999999999	5.56205546865439\\
1.75999999999999	5.50782389952346\\
2.19999999999999	5.7455617258843\\
2.63999999999999	6.9698732338112\\
3.08000000000001	7.23935508057767\\
3.52000000000001	7.36248311964107\\
3.96000000000001	7.13662838254001\\
4.40000000000001	6.99814519847362\\
4.84	6.58072030855868\\
5.28	5.14521936481017\\
5.72	2.26076384400014\\
6.16	2.26076384400014\\
6.59999999999999	3.57585581174666\\
7.03999999999999	3.64307740971785\\
7.47999999999999	3.48708796418583\\
7.91999999999999	3.38786763917301\\
8.36000000000001	3.0980204382727\\
8.80000000000001	3.48608223794719\\
9.24000000000001	3.40251566674905\\
9.68000000000001	2.55298787347238\\
10.12	2.16427738175679\\
10.56	1.72824419094292\\
11	1.97129114103407\\
11.44	2.34953872869986\\
11.88	3.37705383388035\\
12.32	4.98305126507927\\
12.76	5.74727771427371\\
13.2	5.86331911814131\\
13.64	5.67350798875373\\
14.08	5.20533010046987\\
14.52	4.5416808949125\\
14.96	3.91984510649627\\
15.4	3.87550760632362\\
15.84	2.75092510831544\\
16.28	3.03454994513936\\
16.72	4.38898566301731\\
17.16	5.53617576542635\\
17.6	5.96543452470422\\
18.04	5.55847669872301\\
18.48	5.19190727406607\\
18.92	4.67142135940605\\
19.36	4.56282002580545\\
19.8	4.8495934183461\\
20.24	5.17718490287405\\
20.68	6.12582972841983\\
21.12	7.6549555956463\\
21.56	9.07014083735911\\
22	10.0596190467068\\
22.44	11.2524502560015\\
22.88	13.5211821514658\\
23.32	13.5914985293524\\
23.76	12.8833008031517\\
24.2	12.3122783790195\\
24.64	12.2520192080094\\
25.08	13.1569409177238\\
25.52	14.5277405740242\\
25.96	15.5403978501688\\
26.4	16.4905222185489\\
26.84	16.544272866313\\
27.28	15.7712111062868\\
27.72	14.5363238321428\\
28.16	13.9766623740413\\
28.6	13.980973389349\\
29.04	14.0043971810974\\
29.48	14.7366489541073\\
29.92	14.779529814289\\
30.36	12.3599725348527\\
30.8	12.3599725348527\\
31.24	12.731419646202\\
31.68	11.1740514505445\\
32.12	11.1292714355743\\
32.56	11.9666365169872\\
33	11.0086125789625\\
33.44	11.0086125789625\\
33.88	10.8684465340905\\
34.32	10.1937051923934\\
34.76	10.0400587140393\\
35.2	10.0165047882616\\
35.64	9.80629719101969\\
36.52	7.94498377601209\\
36.96	7.00157496615333\\
37.4	7.04464767982728\\
37.84	7.0009854425609\\
38.28	6.55341779107593\\
38.72	6.6666007167772\\
39.16	6.98773117546904\\
39.6	7.08297494221847\\
40.04	7.09087940088139\\
40.48	7.20084188140379\\
40.92	7.50013206897299\\
41.36	7.49741829806277\\
41.8	7.39077569659835\\
42.24	7.23117163098894\\
42.68	6.9632143074559\\
43.12	6.54719871907335\\
43.56	6.25577550488364\\
44	5.85459909480431\\
44.88	5.35800592341707\\
45.32	5.10633771187415\\
45.76	4.59048086433913\\
46.2	3.62101377488824\\
46.64	3.37658881990168\\
47.08	3.37658881990168\\
47.52	3.30251015340858\\
47.96	3.08908504827008\\
48.4	2.77179315578903\\
48.84	2.57807342318196\\
49.28	2.54781363494826\\
49.72	2.77954948781502\\
50.16	3.39230803652617\\
50.6	3.7270737669426\\
51.04	3.46944184487623\\
51.48	3.15958734592851\\
51.92	2.69107443982759\\
52.36	2.46461285214616\\
52.8	2.34812585864819\\
53.24	1.82040588320882\\
53.68	1.64359486651634\\
54.12	1.52618977920653\\
54.56	1.70729631373135\\
55	2.02589837776381\\
55.44	2.02972150635193\\
55.88	1.75088602438765\\
56.32	1.74016793439196\\
56.76	1.80699771826212\\
57.2	1.79555787623906\\
57.64	1.84473405958627\\
58.08	2.05838674117072\\
58.52	2.14215768893106\\
58.96	2.49904978220843\\
59.4	3.05966410179201\\
59.84	3.27874865444608\\
60.28	3.2149955188786\\
60.72	3.1141776808569\\
61.16	3.09429064555971\\
62.04	3.15062652096012\\
62.48	3.19095262576494\\
62.92	3.30230812996481\\
63.36	3.5080212056115\\
63.8	3.77850915377982\\
64.24	4.16773239600579\\
65.12	4.18429785184412\\
65.56	4.62304381009434\\
66.44	4.97353628446371\\
66.88	5.28674429497676\\
67.32	5.41015497641638\\
67.76	5.86495769952808\\
68.2	6.43064846607493\\
68.64	6.50139095072464\\
69.96	7.55757751898815\\
70.4	7.98823193540852\\
70.84	8.06891088865694\\
71.28	7.75733557334897\\
71.72	7.24300260860161\\
72.16	6.52441087105953\\
72.6	6.1005343019765\\
73.04	5.96925461038995\\
73.48	5.9867971100112\\
73.92	6.07452245644251\\
74.36	5.7505484770893\\
74.8	5.12014919196452\\
75.24	4.89367427919225\\
75.68	4.96750811399048\\
76.12	5.2276204110901\\
76.56	7.18247921561431\\
77	8.06126417859275\\
77.44	7.73247116024419\\
77.88	6.93332172841929\\
78.32	7.03418757196886\\
78.76	7.86904071888836\\
79.2	7.17310652909043\\
79.64	7.42538299122702\\
80.08	7.49789604977903\\
80.52	6.85556533146496\\
80.96	6.91151510047436\\
81.4	6.87405530520036\\
81.84	4.71407914937515\\
82.28	4.18295454955589\\
82.72	4.61919719178991\\
83.16	6.12983924935645\\
83.6	7.38939243982256\\
84.04	9.51312950398781\\
84.48	9.51312950398781\\
84.92	8.51983369232644\\
85.36	8.34753256782858\\
85.8	8.78496464343249\\
86.24	9.6585261418638\\
86.68	9.30403838863134\\
87.12	10.0132351425203\\
87.56	10.2792207884817\\
88.44	11.4913986880046\\
88.88	11.755989161013\\
89.32	11.4392027569837\\
89.76	10.1967069680327\\
90.2	9.66124609591029\\
90.64	9.71120390909152\\
91.08	9.52636765236451\\
91.52	9.26454631938228\\
91.96	8.95104787583546\\
92.4	7.92172426938058\\
92.84	7.95437756096922\\
93.28	8.53265574596998\\
93.72	8.57173647082547\\
94.16	8.87258233517403\\
94.6	9.26151596262815\\
95.04	9.27243110554801\\
95.48	7.73913450101963\\
95.92	7.00652195663744\\
96.36	7.31878869535973\\
96.8	9.07018192252178\\
97.24	10.1665835538319\\
97.68	11.8642429073136\\
98.12	13.0355102936897\\
98.56	11.9204134531402\\
99	11.3559869992905\\
99.44	11.2005996557605\\
100.32	11.1824750129614\\
100.76	10.520737442808\\
101.2	10.490471504133\\
101.64	8.92523928931186\\
102.08	8.83574391092804\\
102.52	9.25712482487756\\
102.96	9.51005511782546\\
103.4	10.0963872371017\\
103.84	8.83607606481115\\
104.28	8.14560017895147\\
104.72	7.78872263289099\\
105.16	8.01872387142882\\
105.6	8.40634841310532\\
106.04	8.64707968639365\\
106.48	8.64467889601639\\
107.36	7.70000194546969\\
107.8	7.67989861831802\\
108.24	7.63399880573667\\
108.68	7.41261086124817\\
109.56	7.40760161754213\\
110	7.21828354934209\\
110.44	6.98513578930388\\
110.88	6.80100749828227\\
111.32	6.15421282951473\\
111.76	6.17144872289205\\
112.2	6.40776288705854\\
112.64	6.43933524788559\\
113.08	6.20115857986488\\
113.52	6.22370492351791\\
113.96	6.54410523203131\\
114.4	6.52081081887928\\
114.84	6.14091851558314\\
115.28	6.21175931088985\\
115.72	6.35060768496476\\
116.16	6.22096585314782\\
116.6	6.11172168800968\\
117.04	6.09598851045865\\
117.48	6.01615815833409\\
117.92	6.01744419776759\\
118.36	5.95271039759157\\
118.8	5.76298218197334\\
119.24	5.83389050290177\\
119.68	5.65278120816993\\
120.12	5.58354850054243\\
120.56	5.71374556826117\\
121	5.60373622811107\\
121.44	6.02851128865498\\
121.88	6.24860570262805\\
122.32	6.67524117015651\\
122.76	7.01709146122127\\
123.2	7.64868654256193\\
123.64	7.96120202701439\\
124.08	8.24521022372787\\
124.52	8.86115733534535\\
124.96	9.44749388768381\\
125.84	10.2555738234713\\
126.28	10.4323831573338\\
126.72	10.3616597025767\\
127.16	10.0232031846207\\
127.6	8.94878059567563\\
128.04	7.69964966422668\\
128.48	6.5179569154007\\
128.92	5.48626511605917\\
129.36	5.00514359317131\\
129.8	4.89502662093201\\
130.24	4.9983821190194\\
130.68	5.64675680342069\\
131.56	6.86128340629938\\
132	7.24810645937129\\
132.44	7.37717380177835\\
133.32	7.60399322763118\\
133.76	7.91152373572345\\
134.2	7.88722595971055\\
134.64	7.60906447948435\\
135.08	7.59312206484276\\
135.52	7.85482845698343\\
135.96	7.54185373944637\\
136.4	7.13039671282155\\
136.84	7.16648781815033\\
137.28	8.65359674085494\\
137.72	8.19727836628866\\
138.16	8.17615950122902\\
138.6	8.19540078349684\\
139.04	7.96710347746716\\
139.48	7.41537636753063\\
139.92	7.43203637334099\\
140.36	7.37408472680053\\
140.8	6.96989343831683\\
141.24	6.75446556576742\\
141.68	7.28350630866765\\
142.12	7.27754447905107\\
142.56	6.77411881767921\\
143	6.80574589471817\\
143.44	7.18298720602144\\
143.88	7.16342858267603\\
144.32	6.17991761789352\\
144.76	6.35275255253256\\
145.2	5.97173746195705\\
145.64	5.74460829547834\\
146.08	6.16210555828968\\
146.52	7.2907399968941\\
147.4	10.0476677237685\\
147.84	10.5218764726214\\
148.28	10.0040921581974\\
148.72	9.94089741369953\\
149.16	9.9577254618867\\
149.6	9.16188768766381\\
150.04	7.05697026867125\\
150.48	6.96371581529561\\
150.92	8.27241782090312\\
151.36	9.15822199356944\\
151.8	10.192452219656\\
152.24	13.2265520112368\\
152.68	14.5996883339949\\
153.12	14.3654637263938\\
153.56	13.1285002174889\\
154	11.782920274331\\
154.44	11.6656948434473\\
154.88	12.389236103001\\
155.32	13.2088513775329\\
155.76	13.4588789739105\\
156.2	13.4006244246929\\
156.64	13.4691663462991\\
157.08	13.6254381616621\\
157.52	12.5397817032011\\
157.96	12.434658257569\\
158.4	13.628584255279\\
158.84	14.2696943130359\\
159.28	14.6924317259065\\
159.72	13.2855910626181\\
160.16	12.6020628817068\\
160.6	13.7235416234349\\
161.04	14.9816086539936\\
161.48	15.5050437749703\\
161.92	14.3363799121454\\
162.36	14.0600427937982\\
162.8	13.5948670471898\\
163.24	12.7785278082109\\
164.12	11.3632682858044\\
164.56	10.7196048956999\\
165	10.1416741373041\\
165.88	9.09740678776549\\
166.32	8.64638044144905\\
166.76	8.1378040627593\\
168.08	6.7332095066773\\
168.52	6.2861044253232\\
169.4	5.56101512495701\\
169.84	5.24031936916109\\
170.72	4.49179202919842\\
171.16	4.17378723921072\\
172.48	3.30912003268153\\
173.36	2.78066301961363\\
173.8	2.45166846947842\\
174.24	2.21325284888141\\
174.68	2.07934312216369\\
175.12	1.87675460256642\\
175.56	1.71571192857917\\
176.44	1.47803040891765\\
176.88	1.39460609185639\\
177.32	1.34299449580951\\
177.76	1.32921274593772\\
178.64	1.34100468266007\\
179.08	1.39445571206406\\
179.96	1.47316322268503\\
180.4	1.58317656016487\\
180.84	1.62050384105518\\
181.28	1.70758750899924\\
181.72	1.7624968872521\\
182.16	1.78512726958755\\
183.04	1.79010958934981\\
183.48	1.91880336187236\\
183.92	1.95885042513569\\
184.8	2.08437214333682\\
185.24	2.15059529797938\\
185.68	2.31783577225025\\
186.12	2.51489956304468\\
186.56	3.28167533103712\\
187.44	4.47004202255013\\
187.88	3.89833707942779\\
188.32	1.76699573764512\\
188.76	0.931524947601531\\
189.2	0.901528007702382\\
190.08	1.4977534960473\\
190.52	2.08357440002024\\
190.96	1.83279728061299\\
191.4	1.35971522030587\\
191.84	1.32175646965734\\
192.72	1.47548926935067\\
193.16	1.68036628549834\\
193.6	2.02264993382073\\
194.04	2.85583633919811\\
194.48	3.42248423303175\\
195.36	4.72597436609897\\
195.8	4.72597436609897\\
196.24	4.27871031561995\\
196.68	4.26572503975154\\
197.12	3.02955441062733\\
197.56	2.04940899789321\\
198	2.24209923798608\\
198.44	2.50916507879734\\
198.88	3.2787182922338\\
199.32	3.65433461477289\\
199.76	4.42824266483453\\
200.2	5.69743000683599\\
200.64	6.38957365819655\\
201.08	6.93739739863048\\
201.52	7.39384178941395\\
201.96	7.00361577290408\\
202.4	6.51097650477934\\
202.84	6.09604791417536\\
203.28	4.06187936138127\\
203.72	4.05887348943241\\
204.16	5.74680237177733\\
204.6	6.93625047387047\\
205.04	8.52835262763884\\
205.48	9.19939929888082\\
205.92	9.11582889655071\\
206.36	8.09224971395383\\
206.8	7.2438795549194\\
207.24	7.26555931205377\\
207.68	7.91406101579847\\
208.12	8.27086582459202\\
208.56	8.41477226439829\\
209	8.01521228749834\\
209.44	7.85049102480158\\
209.88	7.17938547738785\\
210.32	6.41538045835318\\
210.76	6.01602625793714\\
211.2	5.67328097722879\\
211.64	5.43165825776154\\
};
\addlegendentry{9D}

\end{axis}
\end{tikzpicture}% 